%% file: uconnect-trpms.tex
\documentclass[lettersize,journal]{IEEEtran}

\input{./preamble}
\input{./acronyms}	
\input{./macros}

\begin{document}
	
	\title{Uconnect: Synergistic Spectral CT Reconstruction with U-Nets Connecting the Energy bins}
	
	\author{Zhihan Wang, Alexandre Bousse, \IEEEmembership{Member,~IEEE},  Franck Vermet, Jacques Froment, B\'eatrice Vedel, Alessandro Perelli,  Jean-Pierre Tasu, Dimitris Visvikis, \IEEEmembership{Fellow, IEEE}
		\thanks{This work did not involve human subjects or animals in its research.}
        \thanks{This work was supported by the French National Research Agency (ANR) under grant No ANR-20-CE45-0020, and by France 2030 framework program, Centre Henri Lebesgue, under grant No ANR-11-LABX-0020-01.}
		\thanks{Zhihan Wang is with Univ. Brest, LaTIM, INSERM, UMR 1101, 29238 Brest, France,  and also with Univ. Brest, CNRS, UMR 6205, LMBA, 29285 Brest, France.}
		\thanks{Alexandre Bousse and Dimitris Visvikis are with Univ. Brest, LaTIM, INSERM, UMR 1101, 29238 Brest, France.}
		\thanks{Franck Vermet is with Univ. Brest, CNRS, UMR 6205, LMBA, 29285 Brest, France.}
		\thanks{Jacques Froment and Béatrice Vedel are with Univ. Bretagne Sud, CNRS, UMR 6205, LMBA, F-56000 Vannes, France.}
		\thanks{Alessandro Perelli is with the School of Science and Engineering, University of Dundee, DD1~4HN~Dundee, U.K.}
		\thanks{Jean-Pierre Tasu is with Univ. Brest, LaTIM, INSERM, UMR 1101, 29238 Brest, France, and also with the Department of Radiology, University Hospital Poitiers, 86000 Poitiers, France.}
		\thanks{Corresponding authors: A. Bousse, \texttt{bousse@univ-brest.fr} }
	}
	
	\markboth{IEEE TRANSACTIONS ON RADIATION AND PLASMA MEDICAL SCIENCES,~Vol.~XX, No.~X, XXX~2023}%
	{Shell \MakeLowercase{\textit{et al.}}: A Sample Article Using IEEEtran.cls for IEEE Journals}
	
	
	\maketitle
	
	\input{./intro}

	\input{./method}

	\input{./results}
	
	\input{./discussion}
	
	\input{./conclusion}

	\section{Acknowledgments}
	
	All authors declare that they have no known conflicts of interest in terms of competing financial interests or personal relationships that could have an influence or are relevant to the work reported in this paper.

	\bibliographystyle{IEEEtran}
	\bibliography{./references}

\end{document}

%% file: preamble.tex
\usepackage{amsmath,amssymb,amsfonts}
\usepackage{bm}
\usepackage{lipsum}
\usepackage{overpic}

\usepackage{algorithmic}
\usepackage[ruled]{algorithm2e}
\SetKwComment{Comment}{/* }{ */}
\usepackage{graphicx}

\usepackage{microtype}
\usepackage{lipsum}
\usepackage{acro}
\acsetup{single}

\usepackage{comment}

\usepackage{cite}

\usepackage{tikz}
\usepackage{tikzscale}
\usetikzlibrary{arrows,shapes,calc,decorations.pathreplacing,intersections,quotes,angles,positioning,fit,petri,through,shadows,plotmarks,spy}
\usepackage{pgfplots}
\usepackage{subfigure}

\usepackage{import}
\subimport{./}{init}

\usetikzlibrary{positioning}
\usetikzlibrary{3d}

\def\ConvReluColor{rgb:yellow,5;red,1.5;white,5}
\def\PoolColor{rgb:red,1;white,0.4}
\def\UnpoolColor{rgb:blue,2;green,1;white,1.3}
\def\ConcatColor{rgb:blue,5;red,2.5;white,5}

\newcommand{\copymidarrow}{\tikz \draw[-Stealth,line width =0.8mm,draw={rgb:blue,4;red,1;green,1;black,3}] (-0.3,0) -- ++(0.3,0);}

\definecolor{aMagenta}{RGB}{255,21,115}
\definecolor{aOrange}{RGB}{255,161,21}
\definecolor{aBlue}{RGB}{16,191,255}

\newlength{\nx}
\newlength{\ny}
\newlength{\tempdima}

%% file: init.tex
\usetikzlibrary{quotes,arrows.meta}
\usetikzlibrary{positioning}

\def\edgecolor{rgb:blue,4;red,1;green,4;black,3}

\usepackage{Ball}
\usepackage{Box}
\usepackage{RightBandedBox}

%% file: acronyms.tex
\DeclareAcronym{PCD}{
	short=PCD,
	long=photon-counting detector,
}

\DeclareAcronym{CT}{
	short=CT, 
	long=computed tomography,
}

\DeclareAcronym{PCCT}{
	short=PCCT, 
	long=photon-counting CT,
}

\DeclareAcronym{DECT}{
	short=DECT, 
	long=dual-energy CT,
}

\DeclareAcronym{FBP}{
	short=FBP,
	long=filtered backprojection,
}

\DeclareAcronym{2D}{
	short=2-D,
	long=2-dimensional,
}

\DeclareAcronym{3D}{
	short=3-D,
	long=3-dimensional,
}

\DeclareAcronym{MRI}{
	short=MRI,
	long=magnetic resonance imaging,
}

\DeclareAcronym{PET}{
	short=PET,
	long=positron emission tomography,
}

\DeclareAcronym{MBIR}{
	short=MBIR, 
	long=model-based iterative reconstruction,
}

\DeclareAcronym{WLS}{
	short=WLS, 
	long=weighted least squares,
}
\DeclareAcronym{PWLS}{
	short=PWLS, 
	long=penalized WLS,
}

\DeclareAcronym{MSE}{
	short=MSE, 
	long=mean squared error,
}

\DeclareAcronym{TV}{
	short=TV, 
	long=total variation,
}

\DeclareAcronym{TGV}{
	short=TGV, 
	long=total generalized variation,
}

\DeclareAcronym{TNV}{
	short=TNV, 
	long=total nuclear variation,
}

\DeclareAcronym{JTV}{
	short=JTV, 
	long=joint TV,
}

\DeclareAcronym{DTV}{
	short=DTV, 
	long=directional TV,
}

\DeclareAcronym{ASSIST}{
	short=ASSIST, 
	long=aided by self-similarity in image-spectral tensors,
}

\DeclareAcronym{RED-CNN}{
	short=RED-CNN, 
	long=residual encoder-decoder convolutional neural network,
}

\DeclareAcronym{PRISM}{
	short=PRISM, 
	long=prior rank intensity and sparsity model,
}

\DeclareAcronym{PICCS}{
	short=PICCS, 
	long=prior image constrained compressed sensing ,
}

\DeclareAcronym{PLS}{
	short=PLS, 
	long=parallel level sets,
}

\DeclareAcronym{DiL}{
	short=DiL, 
	long=dictionary learning,
}

\DeclareAcronym{TDiL}{
	short=TDiL, 
	long=tensor DiL,
}

\DeclareAcronym{CDiL}{
	short=CDiL, 
	long=convolutional DiL,
}

\DeclareAcronym{CNN}{
	short=CNN, 
	long=convolutional neural network,
}

\DeclareAcronym{GAN}{
	short=GAN, 
	long=generative adversarial network,
}

\DeclareAcronym{WGAN}{
	short=W-GAN, 
	long=Wasserstein GAN,
}

\DeclareAcronym{VAE}{
	short=VAE, 
	long=variational autoencoder,
}

\DeclareAcronym{GT}{
	short=GT,
	long=ground truth,
}

\DeclareAcronym{SNR}{
	short=SNR,
	long=signal-to-noise ratio,
}

\DeclareAcronym{PSNR}{
	short=PSNR,
	long=peak signal-to-noise ratio,
}

\DeclareAcronym{SSIM}{
	short=SSIM,
	long=structural similarity index measure,
}

\DeclareAcronym{DL}{
	short=DL,
	long=deep learning,
}

\DeclareAcronym{SQS}{
	short=SQS,
	long=separable quadratic surrogate,
}

\DeclareAcronym{LBFGS}{
	short=L-BFGS,
	long=limited-memory  Broyden-Fletcher-Goldfarb-Shanno,
}

\DeclareAcronym{SDR}{
	short=SDR,
	long=Synergistic Deep Regularization,
}

\DeclareAcronym{XCAT}{
	short=XCAT,
	long=extended cardiac-torso,
}

\DeclareAcronym{HU}{
	short=HU,
	long=Hounsfield unit,
}

\DeclareAcronym{VMI}{
	short=VMI,
	long=virtual mono-energetic image,
}

%% file: macros.tex
\newcommand{\boldb}{\bm{b}}

\newcommand{\boldf}{\bm{f}}

\newcommand{\boldx}{\bm{x}}
\newcommand{\boldy}{\bm{y}}
\newcommand{\boldz}{\bm{z}}

\newcommand{\boldxi}{\bm{\xi}}

\newcommand{\boldA}{\bm{A}}

\newcommand{\boldD}{\bm{D}}

\newcommand{\boldH}{\bm{H}}
\newcommand{\boldI}{\bm{I}}

\newcommand{\boldL}{\bm{L}}

\newcommand{\boldP}{\bm{P}}

\newcommand{\boldS}{\bm{S}}
\newcommand{\boldT}{\bm{T}}

\newcommand{\boldW}{\bm{W}}
\newcommand{\boldX}{\bm{X}}

\newcommand{\calN}{\mathcal{N}}

\newcommand{\rmd}{\mathrm{d}}
\newcommand{\rme}{\mathrm{e}}

\newcommand{\rms}{\mathrm{s}}

\newcommand{\ybar}{\bar{y}}

\newcommand{\xbar}{\bar{x}}

\renewcommand{\hbar}{\bar{h}}

\newcommand{\boldxhat}{\hat{\bm{x}}}
\newcommand{\boldzhat}{\hat{\bm{z}}}

\newcommand{\argmin}{\operatornamewithlimits{arg\,min}}

\newcommand{\R}{\mathbb{R}}

\newcommand{\diag}{\mathrm{diag}}

\newcommand{\ie}{i.e.,}

\newcommand{\transp}{^\top}

\newcommand{\rowname}[1]
{\rotatebox{90}{\makebox[\tempdima][c]{\scriptsize#1}}}

\renewcommand{\th}{$^\mathrm{th}$}

%% file: intro.tex
\begin{abstract}
	Spectral \ac{CT} offers the possibility to reconstruct attenuation images at different energy levels, which can be then used for material decomposition.
	However, traditional methods reconstruct each energy bin individually and are vulnerable to noise. In this paper, we propose a novel synergistic method for spectral \ac{CT} reconstruction, namely Uconnect. It utilizes trained \acp{CNN} to connect the energy bins to a latent image so that the full binned data is used synergistically. We experiment on two types of low-dose data: simulated and real patient data. Qualitative and quantitative analysis show that our proposed Uconnect outperforms state-of-art \ac{MBIR} techniques as well as \ac{CNN}-based denoising.
\end{abstract}

\begin{IEEEkeywords}
	Spectral \ac{CT}, Deep learning,  Synergistic Reconstruction, Regularization
\end{IEEEkeywords}

\section{Introduction}\label{sec:introduction}

\IEEEPARstart{S}{pectral} \ac{CT} is fast becoming an important technology in medical imaging for its potential to differentiate between different types of tissues and structures based on their unique spectral properties \cite{hsieh2020photon}. Its applications include lesion detection \cite{lv2012spectral}, material decomposition \cite{long2014multi, wu2020image}, automated bone removal \cite{morhard2009cervical}, etc. As an emerging technology to implement \ac{PCCT} imaging, \acp{PCD} are capable of collecting multiple sets of measured data with different spectral information in a single exposure by distinguishing photon energies during data acquisition. On the other hand, with the increasing number of energy bins, the number of X-ray photons in each energy bin is limited, which reduces \ac{SNR} in spectral projections acquired by \acp{PCD}. At the same time efforts to reduce radiation dose, largely motivated by the introduction of potential clinical applications such as lung cancer screening, are associated with decreasing number of projection angles and/or reduced X-ray source flux. In this situation, traditional analytic reconstruction methods such as \ac{FBP} will cause severe artifacts and noise. Therefore, many studies have focused on developing methods to improve the reconstruction quality.

Over the past few years, several sparsity-exploiting methods have been applied to low-dose \ac{CT} reconstruction, such as \ac{TV} \cite{rudin1992nonlinear, zhang2005total, sidky2008image}, \ac{DiL} \cite{xu2012low}, and \ac{TGV} \cite{niu2014sparse}. These methods can be extended to \ac{PCCT} multi-energy reconstruction by applying the sparse regularization to each energy bin separately. However, this kind of strategy ignores dependencies between images acquired at different energy bins, which may lead to suboptimal results. Many algorithms, thus, have been proposed to take advantage of the similarities between different energy bins.

Synergistic regularization terms have been designed to promote structural similarities across channels. Examples include \ac{JTV} \cite{sapiro1996anisotropic,blomgren1998color} which promotes joint sparsity of the gradients of images, \ac{TNV}  \cite{lefkimmiatis2013convex,rigie2015joint} which encourages gradient-sparse solutions in the same way as the conventional \ac{TV} and also favors solutions that have common edge directions in all channels, and \ac{PLS} \cite{ehrhardt2014joint} which promotes images with similar contours. In addition, the low-rank property is often used as a prior assumption to exploit the correlation among different energy bins. Specifically, the spectral \ac{CT} data can be represented as a low-rank matrix, where each row corresponds to an energy bin and each column corresponds to a voxel in the image. In this way, it is possible to obtain more accurate and robust spectral \ac{CT} images during reconstruction \cite{gao2011multi,semerci2014tensor,niu2018nonlocal,xia2019spectral}. Another class of approach consists in enforcing structural similarities with a reference clean image \cite{cueva2021synergistic,zhang2016spectral,yao2019multi,wang2020low}.

Synergistic regularization can also be trained as opposed to the use of a fixed analytical model. \Ac{TDiL} \cite{zhang2016tensor,wu2018low,li2022tensor}  has also emerged as a promising technique for spectral \ac{CT} reconstruction. It is a powerful extension of traditional dictionary learning, where instead of representing the data using a single dictionary, a collection of dictionaries is learned, such that information can be conveyed across channels. However, they rely on the patch-based sparse representation that may be suboptimal as the trained atoms are often shifted version of each other. To remedy this problem, multichannel convolutional \ac{DiL} techniques can be deployed \cite{perelli2022multi}. More details can be found in a recent review \cite{bousse2023systematic}.

Recently, there has been growing interest in applying \ac{DL} techniques to \ac{CT} reconstruction \cite{wang2020deep,ahishakiye2021survey,xia2023physics}. Several studies have shown promising results using various \ac{DL} architectures, such as \acp{CNN}, \acp{GAN}, autoencoders, and unrolling architectures. However, few investigations have been carried out in the domain of spectral \ac{CT}. For example, \cite{mustafa2020sparse,wu2021deep} have proposed \ac{DL}-based spectral \ac{CT} multichannel  post-processing techniques. Recently, an unrolling architecture, namely SOUL-Net \cite{chen2022soul}, was proposed. However the training of such models is challenging, especially with a large number of channels.

In this paper, we propose a novel spectral \ac{CT} reconstruction technique, namely Uconnect, where images at each energy bin are connected to a reference image by a collection of U-Nets. We considered the case of \ac{PCCT} with 6 energy bins, although our method can also be applied to dual-energy \ac{CT}. The training is performed on an image basis, and therefore does not require a computationally expensive supervised training for sinogram-to-image mappings such as unrolling architectures. The U-Nets are pretrained on a separate dataset to map the attenuation image at a reference energy to other energies and are incorporated into a plug-and-play penalty term which can be used in any \ac{MBIR} technique, for any system geometry. To the best of our knowledge, this is the first method that employs a deep-learned penalty for multi-energy \ac{CT} synergistic \ac{MBIR}. This work follows previous work we presented at the 2022 IEEE Medical Imaging Conference \cite{wang2022synergistic}. 

The rest of the paper is structured as follows. Section~\ref{sec:method} briefly reviews the physical model and several existing approaches for spectral \ac{CT} reconstruction, and presents the details on our proposed approach (Uconnect). Section~\ref{sec:experiments} shows the details of experiments on synthetic data and real patient data, as well as qualitative and quantitative comparison results with other approaches. Section~\ref{sec:discussion} discusses several technical issues, while Section~\ref{sec:conclusion} provides our concluding remarks.

%% file: method.tex
\section{Methods}\label{sec:method}

\subsection{Forward Model}

In this work we consider a standard \ac{PCCT} set-up. We consider \ac{2D} images, although the proposed method can generalize to the \ac{3D} case.

The discrete anatomical image takes the form of an energy-dependent vector $\boldx(E) = \left[x_1(E),\dots,x_J(E)\right]\transp \in \R^J$, where for all $j=1,\dots,J$, $x_j(E)$ is the linear attenuation at pixel $j$ and energy $E$, $J$ is the total number of pixels (or voxels) in the image, and `$\transp$' denotes the matrix transposition.

The spectral \ac{CT} system performs measurements along a collection of $I$ rays (from the source to the detectors), with $I=N_\rmd \times N_\rms$, $N_\rmd$ and $N_\rms$ being, respectively, the number of detectors and the number of source positions. The measurements are regrouped into $K$ energy bins, each bin $k$ corresponding to an interval $\left[E_{k-1},E_k\right]$ with $E_0<E_1<\dots<E_K$. For each bin $k=1,\dots, K$ and each ray $i=1,\dots, I$, the expected number of detected photons is given by the Beer-Lambert law as   
\begin{equation}\label{eq:beer-lambert_binned}
	\ybar_{i,k}(\boldx) = \int_0^{+\infty} h_{k}\left(E\right) \cdot \rme^{ -\left[\boldA \boldx(E)\right]_i} \, \rmd E
\end{equation}
where $h_{k}\colon \R^+\to\R^+$ is the photon flux X-ray intensity associated to energy bin $k$. For simplicity we ignored background events such as scatter but they can be incorporated in the model without affecting the rational of the proposed approach. The matrix $\boldA \in \R^{I\times J}$ is the system matrix modeling line integrals along each beam in the discrete case, and for ray $i$ and pixel $j$, $[\boldA]_{i,j}$ is the contribution of pixel $j$ to beam $i$.

Instead of reconstructing $\boldx(E)$ as a function of $E$, we reconstruct a multichannel image $\left\{\boldx_k\right\} \in\left(\R_+^J\right)^K$ where for each $k=1,\dots,K$  the image $\boldx_k$ is an ``average'' attenuation image corresponding to energy bin $k$, using the simplified forward model 
\begin{equation}\label{eq:beer-lambert_discrete}
    \ybar_{i,k}\left(\boldx_k\right) =  \hbar_{k} \cdot \rme^{ -\left[\boldA \boldx_k\right]_i   }  
\end{equation}	
where $\hbar_{k}=\int h_{k}\left(E\right)\rmd E$ is the mean photon flux X-ray intensity for energy bin $k$. The number of detected photons at each bin $k$ and ray $i$ is modeled as a Poisson random variable denoted $y_{i,k}$, \ie{}
\begin{equation}\label{eq:poisson2}
    y_{i,k} \sim \mathrm{Poisson} \left(\ybar_{i,k}\left(\boldx_k\right) \right) \, ,
\end{equation}
and for all $i,i',k,k'$, $y_{i,k}$ and $y_{i',k'}$ are independent if $(i,k)\ne \left(i',k'\right)$. We denote by $\boldy_k = \left[y_{1,k},\dots,y_{I,k}\right]\transp \in \R^I$ the vector of measurement at bin $k$.

\subsection{Conventional Reconstruction}

Each $\boldx_k$ can be reconstructed by minimizing the negative log-likelihood of the measurement $\boldy_k$. To simplify the reconstruction, we approximate the negative log-likelihood with a \ac{WLS} loss function $L_k$ as proposed in \cite{elbakri2002statistical}: 
\begin{align}
	L_k \left(\boldx_k\right) = {} & \sum_{i=1}^I \frac{1}{2}y_{i,k} \left([\boldA \boldx_k]_i - b_{i,k}\right)^2 \\ 
= {} & \frac{1}{2}\left\| \boldA \boldx_k - \boldb_k  \right\|^2_{\boldW_k}	\label{eq:wls}
\end{align}
with $\boldb_k = \left[b_{1,k},\dots,b_{I,k}\right]\transp$, $b_{i,k} = \log \left(\hbar_{k} / y_{i,k}\right)$, 
where we assumed $y_{i,k} >0 $ for all $(i,k)$, and $\boldW_k = \mathrm{diag}\left\{\boldy_k\right\}$ is a diagonal matrix of statistical weights. Finally, each $\boldx_k$ can be reconstructed by \ac{PWLS} with positivity constraint:
\begin{equation}\label{eq:pwls}
	\boldxhat_k = \argmin_{\boldx_k \in \R^J_+}\, L_k\left(\boldx_k\right) + \beta R\left(\boldx_k\right)
\end{equation}
where $\beta>0$ is a weight and $R$ is an edge-preserving regularization term. A popular choice is the Huber regularizer which enforces piecewise smoothness \cite{huber2011robust}, defined as
\begin{align}\label{eq:huber}
	R_\mathrm{H}(\boldx) = \sum_{j=1}^J \sum_{l\in\calN_j} \omega_{j,l}\psi_\delta\left(x_j-x_l\right), \\ 
	\psi_\delta(t) = \left\{  
	\begin{array}{ll}
		\frac{1}{2}t^2 & \text{if $|t|\le\delta$}  \\
		\delta|t| - \frac{1}{2}\delta^2 & \text{otherwise}  
	\end{array}	
	\right.\nonumber
\end{align}
where $\delta>0$, $\calN_j$ is the index set of the neighbor of pixel $j$, and $\omega_{j,l}$ is the inverse Euclidean distance between $j$ and $l$.

\subsection{Synergistic Penalties: Existing Approaches}\label{sec:syn_pen}

Instead of reconstructing each $\boldx_k$ independently following \eqref{eq:pwls}, the entire collection of images $\left\{\boldx_k\right\}$ can be reconstructed simultaneously in a single optimization problem as 
\begin{equation}\label{eq:synergestic_pwls}
	\{\boldxhat_k\} = \argmin_{\{\boldx_k\} \in \left(\R^J_+\right)^K}  \, \sum_{k=1}^K L_k(\boldx_k) + \beta R_{\mathrm{syn}}\left(   \left\{\boldx_k\right\}\right)
\end{equation}
where in this case $R_{\mathrm{syn}}$ is a synergistic regularization term for the multichannel  image $\left\{\boldx_k\right\}$. Note that the sum of the $L_k$s can be weighted in order to equally smooth the $\boldx_k$s. In this section we describe existing categories of methods for spectral~\ac{CT} reconstruction and we highlight the methods we implemented for comparison with our proposed approach outlined in the following section~\ref{sec:proposed}.

A first category of methods consists in using a regularizer that enforces structural similarities between channels. An example is \ac{JTV}, whose regularizer is defined as 
\begin{equation}\label{eq:jtv}
	R_\mathrm{JTV}\left(   \left\{\boldx_k\right\}\right) = \sum_{j=1}^J \sqrt{  \sum_{k=1}^K   \left\|  \left[\nabla \boldx_k\right]_j   \right\|^2_2           }
\end{equation}
where $\nabla \colon \R^J \to \left(\R^J\right)^2$ denotes the discrete gradient. This promotes joint sparsity of the gradient of the $\boldx_k$s. Other examples include \ac{TNV} \cite{lefkimmiatis2013convex,rigie2015joint} and \ac{PLS} \cite{ehrhardt2014joint}.

A second category involves controlling the rank of the multichannel image to promote linear dependencies between the channels. It is reasonable to assume some form of ``low rankness'' due to the fact that the energy dependency of human tissues can be expressed as a linear combination of only two materials. In \cite{gao2011robust}, a low-rank and sparse (LRS) decomposition was proposed for the multichannel image matrix $\boldX = [\boldx_1,\dots,\boldx_K] \in \R^{J \times K}$ such that $\boldX = \boldL + \boldS$, where the low-rank matrix $\boldL$ represents the underlying structure or common information between channels and the sparse matrix $\boldS$ reflects distinct spectral features across channels. A synergistic penalty is then defined as
\begin{equation}\label{eq:RPCA-4DCT}
	R_{\mathrm{LRS}}\left(\boldX\right) = \gamma\left\| \boldL \right\|_* + \left\|\boldT\boldS \right\|_1
\end{equation}
where $\boldT$ is a sparsifying transform (such as the gradient or a wavelet transform) with $\gamma>0$ and the nuclear norm $\|\cdot\|_*$ is a convex relaxation of the rank for an efficient optimization. The method \ac{ASSIST} in \cite{xia2019spectral} is a generalization of this approach that selects the $M$ most similar $d$-dimensional patches within a search window of a clean reference (or prior) image $\boldx^\mathrm{prior}$, to form a collection of $d\times M \times K$ tensor units of which the nuclear norm is computed.

A third category of methods consists in reconstructing each channel $\boldx_k$ independently by solving \eqref{eq:pwls} with a single-channel penalty that enforces similarities with a reference low-noise prior image $\boldx^\mathrm{prior}\in\R^J$ \cite{cueva2021synergistic,zhang2016spectral,yao2019multi,wang2020low}. For example in \cite{cueva2021synergistic} the \ac{DTV}  regularizer  promotes common edge directions with the prior image, and is defined for all $\boldx\in\R^J$ as
\begin{equation}\label{eq:dtv}
	R_\mathrm{DTV}(\boldx) = \sum_{j=1}^J  \left\| \bm{\Pi}_j [\nabla \boldx]_j \right\|_2
\end{equation}
where $\bm{\Pi}_j = \boldI - \bm{\xi}_j \otimes \bm{\xi}_j$, $\boldI$ being the $2\times2$ identity matrix and $\otimes$ being the outer product of vectors, and 
\begin{equation}\label{eq:dtv_xi}
	\boldxi_j = \eta  \frac{[\nabla \boldx^\mathrm{prior}]_j}{ \sqrt{ \left\|[\nabla \boldx^\mathrm{prior}]_j\right\|_2^2+\epsilon}}
\end{equation}
with $\epsilon>0$ avoiding singularities when $[\nabla \boldx^\mathrm{prior}]_j = 0$, and $\eta<1$ depending on how much influence $\boldx^\mathrm{prior}$ has. In particular, we observe that \cite{bungert2018blind}
\begin{equation}
	\left(1-\eta^2\right) R_\mathrm{TV} \le R_\mathrm{DTV} \le R_\mathrm{TV}
\end{equation}
where the \ac{TV} regulariser $R_\mathrm{TV}$ corresponds to $R_\mathrm{DTV}$ with $\bm{\Pi}_j = \boldI_{\R^2}$ for all $j$, and the lower bound is reached when $[\nabla \boldx]_j$ and $\boldxi_j$ are collinear for all $j$. $R_\mathrm{DTV}(\boldx)$ is minimized when the gradient vanishes and therefore potential artifacts in $\boldx^\mathrm{prior}$ do not propagate in the reconstructed image. Similar approaches include \cite{chen2008prior,yu2016spectral,wang2020low} which use the gradient of the difference between each $\boldx_k$ and $\boldx^\mathrm{prior}$. The choices of $\boldx^\mathrm{prior}$ will be discussed in Section~\ref{sec:experiments}.

\subsection{Proposed Learned Synergistic Penalty: Uconnect}\label{sec:proposed}

The traditional regularization techniques presented in Section~\ref{sec:syn_pen} enforce a fixed handcrafted penalty on the reconstructed image due to certain assumptions about its structure, such as sparsity or smoothness. However, these assumptions may not always be valid in practice. By contrast, learned penalty functions adjust adaptively according to the data's specific characteristics, providing more accurate and adaptable reconstructions.

\subsubsection{General principle}\label{sec:model}

Learned synergistic penalties can be derived from multichannel \ac{DiL}, which consists in training a collection of dictionaries $\boldD_1,\dots,\boldD_K$, i.e., over-complete bases comprising $S$ atoms, to approximately represent each image $\boldx_k$ with a fraction of their columns with a single sparse code $\boldz\in\R^S$, that is to say, 
$\boldx_k \approx \boldD_k\boldz$ for all $k$, in order to convey the information across channels.  Unfortunately,  a large number of atoms is required  to accurately represent all possible  spectral images, which can compromise training. Therefore,  to reduce the complexity, the image is  split into $P$ smaller $d$-dimensional ``patches''  with $d\ll J$. The synergistic penalty is therefore defined as follows:
\begin{align}
    R_{\mathrm{dict}} \left( \{\boldx_k\}  \right) = {} & \min_{\{\boldz_p\}\in (\R^{S})^P} \, \sum_{k=1}^K\sum_{p=1}^P\frac{1}{2}  \|\boldP_p\boldx_k - \boldD_k \boldz_p \|^2_2 \nonumber  \\
    & + \alpha \|\boldz_p \|_q \label{eq:dic_learning}
\end{align}
where $\boldP_p \in \R^{d\times J}$ is the $p$\th{} patch extractor,    $\boldD_k \in \R^{d\times S}$ is the $k$\th{} trained dictionary, $\boldz_p \in \R^S$ is the sparse coding (same for all channel $k$), $\|\cdot\|_q$ is either the $\ell_1$-norm or the $\ell_0$ semi-norm, and $\alpha>0$. In \cite{zhang2016tensor} the authors used tensor dictionaries to represent the spectral image based on canonical polyadic decomposition, which is similar to \eqref{eq:dic_learning}. Similar coupled dictionary models have been used in multimodal imaging synergistic reconstruction, for example in \ac{PET}/\ac{MRI} \cite{sudarshan2018joint,sudarshan2020joint} and multi-contrast \ac{MRI} \cite{song2019coupled}.

Despite the possibility to extract patches at random locations over the image, patch-based \ac{DiL} may suffer from inefficiency due to the shift-variant nature of the atoms that can generate duplicates during training. Furthermore, the utilization of many neighboring/overlapping patches across the training images is not optimal for sparse representation, as sparsification is performed independently on each patch.

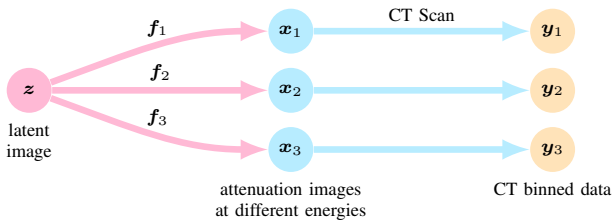
\begin{figure}[htbp]
	\centering
	\begin{tikzpicture}
    	\scriptsize
    	\node (z) at (0,0)  [rounded corners,align=center,fill=aMagenta!30,circle,minimum size=17pt] {$\boldz$} ;	
    	\node (mu1) at ([xshift=3\nx,yshift=0.8\ny]z)  [rounded corners,align=center,fill=aBlue!30,circle,minimum size=17pt] {$\boldx_1$} ;	
    	\node (mu2) at ([xshift=3\nx,yshift=0\ny]z)  [rounded corners,align=center,fill=aBlue!30,circle,minimum size=17pt]{$\boldx_2$} ;
    	\node (mu3) at ([xshift=3\nx,yshift=-0.8\ny]z)  [rounded corners,align=center,fill=aBlue!30,circle,minimum size=17pt]
    	{$\boldx_3$} ;

    	\node (y1) at ([xshift=3\nx,yshift=0\ny]mu1)  [rounded corners,align=center,fill=aOrange!30,circle,minimum size=17pt] {$\boldy_1$} ;	
    	\node (y2) at ([xshift=3\nx,yshift=0\ny]mu2)  [rounded corners,align=center,fill=aOrange!30,circle,minimum size=17pt]{$\boldy_2$} ;
    	\node (y3) at ([xshift=3\nx,yshift=0\ny]mu3)  [rounded corners,align=center,fill=aOrange!30,circle,minimum size=17pt]
    	{$\boldy_3$} ;
    	
    	\node (zz) at ([xshift=0\nx,yshift=-0.7\ny]z)  [align=center] {latent \\  image} ;	
    	\node (mu) at ([xshift=0\nx,yshift=-0.7\ny]mu3)  [align=center] {attenuation   images \\ at different energies} ;	
    	\node (y) at ([xshift=0\nx,yshift=-0.53\ny]y3)  [align=center] {CT binned data} ;	
    	
    	\draw[>=latex,->,out=20,in=180,line width=2.5pt,aMagenta!30] (z.20) to node[midway,above,align=center] {$\textcolor{black}{\boldf_1}$} (mu1.180) ;
    	\draw[>=latex,->,out=0,in=180,line width=2.5pt,aMagenta!30] (z.0) to node[midway,above,align=center] {$\textcolor{black}{\boldf_2}$} (mu2.180) ;
    	\draw[>=latex,->,out=-20,in=180,line width=2.5pt,aMagenta!30] (z.-20) to node[midway,above,align=center] {$\textcolor{black}{\boldf_3}$} (mu3.180) ;

    	\draw[>=latex,->,out=0,in=180,line width=2.5pt,aBlue!30] (mu1.0) to node[midway,above,align=center] {\textcolor{black}{\ac{CT} Scan}} (y1.180) ;
    	\draw[>=latex,->,out=0,in=180,line width=2.5pt,aBlue!30] (mu2.0) to node[midway,above,align=center] {} (y2.180) ;
    	\draw[>=latex,->,out=0,in=180,line width=2.5pt,aBlue!30] (mu3.0) to node[midway,above,align=center] {} (y3.180) ;

	\end{tikzpicture}	
	\caption{Schematic representation of our generative model with $K=3$ energy bins.}\label{fig:gen_model}
\end{figure}

In this work we propose an alternative to patch-based \ac{DiL}. We assume that each image $\boldx_k$ is acquired from a single object $\boldz \in \R^N$ (non necessarily sparse), $N$ being the dimension of the latent space (in \ac{DiL} $N$ corresponds to the number of atoms $S$), with a collection of trained \acp{CNN} $\left\{\boldf_k \right\}$, such that
\begin{equation}\label{eq:model}
\boldf_k(\boldz) \approx \boldx_k, \quad \forall k=1,\dots,K \, .
\end{equation}
The mappings $\left\{\boldf_k \right\}$ play the role of the ``non-linear'' dictionaries for the entire image (without patches), which connect all energy bins through $\boldz$. The proposed joint regularization term is therefore defined as
\begin{equation}\label{eq:r-new}
    R_\mathrm{NN}\left( \{\boldx_k\}  \right)  =   \min_{\boldz \in \R^N} \, \sum_{k=1}^K \frac{1}{2} \gamma_k\left\| \boldf_k(\boldz) - \boldx_k \right\|^2_2   
    +  \alpha H(\boldz)
\end{equation}
where $H$ is a penalty term on $\boldz$. This representation is a generalization of \eqref{eq:dic_learning} with non-linear mappings $\left\{\boldf_k \right\}$ instead of dictionaries, and without patch extraction.  The regularizer $R_\mathrm{NN}\left(\left\{\boldx_k\right\}\right)$ is small if (i) each image $\boldx_k$ is close to  $\boldf_k(\boldz)$ for some $\boldz$ and (ii) $\boldz$ is smooth (in the sense of $H$). By doing so, the $\boldx_k$s are ``connected'' via a single latent variable $\boldz$, (cf. the schematic representation of the generative model in Fig.\@~\ref{fig:gen_model}) such that each $\boldx_k$ is reconstructed using the entire measurement data at each energy bin. 

The role of the $\gamma_k$s is to give more weight to less noisy images. A logical choice would be $\gamma_k =  \|\boldy_k\|_1 / \sum_k \|\boldy_k\|_1 $. However, most of the counts come from the rays that do not intersect the patient, and hence this approach does not account for the loss of counts due to higher attenuation. We therefore define $\gamma_k =  \|\check{\boldy}_k\|_1 / \sum_k \|\check{\boldy}_k\|_1  $ through manual experimentation, where $\check{\boldy}_k\in\R^{I'}$, $I'<I$, is the vector composed of the smallest first quarter of $\boldy_k$ in order to only account for rays intersecting the patient.

The training of the $\boldf_k$s can be unsupervised, for example using \acp{WGAN} \cite{arjovsky2017wasserstein} or \acp{VAE} \cite{duff2021regularising} with some latent variable $\boldz$ that has no physical interpretation. However, this training can be challenging for a large number of channels $K$ as the $\boldf_k$s must be jointly trained. We therefore propose the following alternative. Our method, namely \emph{Uconnect}, consists of ``connecting'' the energies by the mean of image-to-image U-Nets \cite{ronneberger2015u} $\boldf_k \colon \R^J\to \R^J$ trained in a supervised fashion to map a \ac{CT} image at energy bin $1$ to energy bin  $k$, that is to say\footnote{An alternative to $R_{\mathrm{NN}}$ consists in dropping the latent variable $\boldz$, i.e., $R_\mathrm{NN}^\mathrm{alt} \left( \{\boldx_k\}  \right)  =    \sum_{k=2}^K \frac{1}{2} \gamma_k\left\| \boldf_k(\boldx_1) - \boldx_k \right\|^2_2   
	+  \alpha H(\boldx_1)$. Note that this penalty is limited image-to-image \acp{CNN} while $R_\mathrm{NN}$ in \eqref{eq:r-new} can use any generative model defined on a latent space.},
\begin{equation}\label{eq:mapping}
	\boldf_k(\boldx_1) \approx  \boldx_k , \quad \forall k=1,\dots,K \, ,
\end{equation}
In this model, the latent variable is an image, and the dimension of the latent space is $N=J$. This approach is somehow similar to the \ac{CNN} representation proposed in \cite{gong2018iterative} for \ac{PET}, in the sense that they both  assume that the images are outputs of some \acp{CNN} trained with clean images as target in order to  reduce the noise.  However, our method utilizes a single input $\boldz$ for all images, thus further reducing the noise as we will show in Section~\ref{sec:pre-eval-xcat} and Section~\ref{sec:pre-eval-real}. In addition, the generative model is used within a penalty term, which is less constraining than the strict equality $\boldx = \boldf(\boldz)$ used in \cite{gong2018iterative}.

The choice of the reference image $\boldz \approx \boldx_1$, \ie{} the lowest-energy bin, is not arbitrary. It is generally accepted that \ac{CT} images of the same anatomical part but acquired at different energies exhibit similar patterns despite different attenuation values. To take advantage of this similarity, we designated the \ac{CT} image obtained from the lowest-energy bin as the reference image owing to the fact that low-energy attenuation images contain more features than high-energy ones, therefore high-energy to low-energy mappings would have to be trained to reveal features, which is more challenging than the other way around. By leveraging this approach, the shared characteristics across energy levels can enhance the accuracy and quality of spectral \ac{CT} reconstructions. Note that the model \eqref{eq:mapping} does not replicate the exact physics, as the attenuation at each energy cannot be predicted from a single image; the material decomposition should be used to generate the \acp{VMI}, followed by an averaging over each energy bin. However, as we will show in Section~\ref{sec:experiments}, the lowest-energy image can reasonably predict the images at other energies. Similar observation was made in \cite{liu2021generation}.

To finish, $\boldz$ is an image and therefore the penalty function $H$  can be any image regularizer such as $R_\mathrm{H}$ \eqref{eq:huber}, or even $R_\mathrm{DTV}$ \eqref{eq:dtv}. 

\subsubsection{Training and Implementation}\label{sec:TrainingUnet}

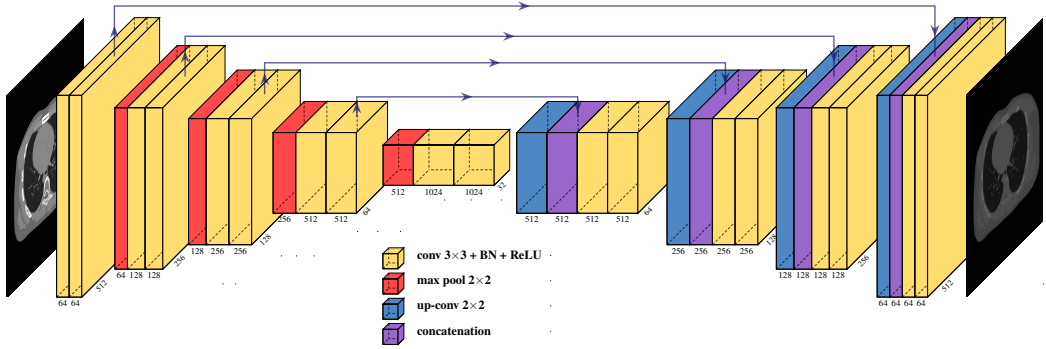
\begin{figure*}[htbp]
	\centering
	\resizebox{\linewidth}{!}{
		\centering
		\input{./unet}
	}
	\caption{U-Net architecture used for the $\boldf_k$.}\label{fig:unet}
\end{figure*}

In this work, $K$ mappings $\left\{\boldf_k \right\}$ need to be trained for each energy bin $k=1,\dots,K$. The overall network architecture of the U-Net structure we consider for each $\boldf_k$ is shown in Fig.~\ref{fig:unet}. This \ac{CNN} has 23 convolutional layers in total. The number of trainable parameters is around 30 millions. 
The training is performed using a collection of $L$  spectral \ac{CT} images $\left\{\boldx^\mathrm{train}_{k,l}\right\}_{k,l=1}^{K,L}$, $\boldx^\mathrm{train}_{k,l}$ representing the \ac{CT} image at bin $k$ from the $l$\th{} dataset. Since the attenuation values vary across energy bins we introduced scaling factors $s_{k,l}$ in order to facilitate training, such that
\begin{equation}\label{eq:training}
	\boldf_k \left(\boldx^\mathrm{train}_{1,l}\right) \approx s_{k,l} \boldx^\mathrm{train}_{k,l}, \quad \forall k=1,\dots,K, \quad \forall l=1,\dots,L
\end{equation}
with $s_{k,l} = \frac{ \|\boldx^\mathrm{train}_{1,l}\|_1}{\|\boldx^\mathrm{train}_{k,l}\|_1}$. We consider $\left\{\boldx^\mathrm{train}_{1,l}\right\}_{l=1}^{L}$ as input images and $\left\{\boldx^\mathrm{train}_{k,l}\right\}_{k,l=1}^{K,L}$ as their corresponding target images. $L$ is the amount of labeled data for training. Note that we also train the network for $k=1$, i.e., bin 1 to bin 1, in order to take advantage of the denoising properties of the $\boldf_k$. The training was achieved with  \ac{MSE} and Adam optimizer. 

To account for the scaling factors, a modified version of the synergistic penalty \eqref{eq:r-new} is used for our experiments as follows:
\begin{equation}\label{eq:r-final}
	R_{\mathrm{NN}}\left(\left\{\boldx_k\right\}\right) =  \argmin_{\boldz \in \R^N} \, \sum_{k=1}^K \frac{1}{2} \gamma_k\left\| \boldf_k(\boldz) - \hat{s}_k\boldx_k \right\|^2_2   
{} +  \alpha H(\boldz)
\end{equation}
where $\hat{s}_k$ is an estimated scaling factor, which can be obtained from scout reconstructions, for example by solving \eqref{eq:pwls} for each bin independently. Note that this factor is object-dependent, although we treat it here as a constant.

\subsubsection{Reconstruction Algorithm}\label{sec:algorithm}

Solving \eqref{eq:synergestic_pwls} using the synergistic penalty \eqref{eq:r-final} is performed by alternating minimization with respect to $\boldz$ and $\{\boldx_k\}$, \ie{} given the estimates $\left\{\boldx_k^{(n)}\right\}$ and $\boldz^{(n)}$ at iteration $n$, the new estimates at iteration $n+1$ are given by
\begin{align}
    \boldz^{(n+1)} = {} & \argmin_{\boldz\in\R^J_+}\,  \sum_{k=1}^K \frac{1}{2} \gamma_k \left \| \boldf_k(\boldz) - \hat{s}_k \boldx_k^{(n)} \right \|^2_2  +  \alpha H(\boldz) \label{eq:update_z}  \\
    \boldx_k^{(n+1)} = {} & \argmin_{\boldx\in\R^J_+}  \, L_k(\boldx) + \beta \gamma_k \frac{1}{2} \left\| \boldf_k\left(\boldz^{(n+1)}\right) - \hat{s}_k \boldx \right\|_2^2   \, , \nonumber \\ & \forall k=1, \dots, K  \label{eq:update_x}
\end{align}
We treat each sub-problem with iterative algorithms.

When $H$ is differentiable (e.g., $H=R_\mathrm{H}$), the $\boldz$-update \eqref{eq:update_z} is achieved with a \ac{LBFGS} algorithm \cite{zhu1997algorithm} (initialized from $\boldz^{(n)}$ ) with non-negativity constraints, which is a quasi-Newton optimization algorithm that only requires the gradient of the objective function. We used the \texttt{GradientTape} function in TensorFlow  to compute the gradient of a function with respect to its input variables. When $H$ is non-differentiable (e.g., \ac{TV}), the  minimization can achieved with a Chambolle-Pock algorithm \cite{chambolle2011first}, combined with \ac{LBFGS} for the proximal of $\left\| \boldf_k(\cdot) - \hat{s}_k \boldx_k^{(n)} \right\|^2_2$.

For the $\boldx_k$-update \eqref{eq:update_x}, we used a \ac{SQS} algorithm \cite{elbakri2002statistical}, i.e., using a diagonal matrix $\boldH_k$ that majorizes $\boldA\transp\boldW_k \boldA$ (e.g., $\boldH_k= \diag(\boldA\transp \boldW_k \boldA \bm{1} )$ where $\bm{1}$ is a vector containing $1$ only, see \cite{lange2000optimization}). Let $\boldx_k^{(n,m)}$ be the current estimate at iteration $n$, sub-iteration $m$.
The new image at sub-iteration $m+1$ is given by
\begin{align}
	\tilde{\boldx}_k = {} & \boldx_k^{(n,m)} - \boldH_k^{-1} \boldA\transp \boldW_k \left(\boldA \boldx_k^{(n,m)}  - \boldb_k\right) \nonumber \\
	\boldx_k^{(n,m+1)} = {} &  \left[\frac{\boldH_k \tilde{\boldx}_k  + \beta \gamma_k \hat{s}_k \boldf_k\left(\boldz^{(n+1)}\right)}{\boldH_k + \beta \gamma_k \hat{s}_k^2} \right]_+ \label{eq:x-udpate}
\end{align}
where $[\cdot]_+$ denotes the positive part of a vector.

We summarize the reconstruction steps in Algorithm~\ref{alg:recon}. Before the main loop that alternates minimization with respect to $\boldz$ and $\{\boldx_k\}$, we initialize $\left\{\boldx_k^{(0)}\right\}$ either by images of value $0$ or by preliminary images $\left\{\boldxhat_k\right\}$ reconstructed individually using \ac{WLS} without penalty for faster convergence. Similarly, $\boldz^{(0)}$ can be initialized by $\boldx_1^{(0)}$. The convergence speed of the algorithm depends on $\beta$. Since $R_\mathrm{NN}$ is non-convex, proving the convergence is challenging. However, we observed in our experiments that the algorithm converges, even for large values of $\beta$.

\begin{algorithm}[hbpt]
	\footnotesize
	\caption{Algorithm for multi-energy \ac{CT} reconstruction with Uconnect}
	\label{alg:recon}
	\KwIn{System matrix $\boldA$; Measurements $\left\{\boldy_k\right\}$; Initial images $\left\{\boldx_k^{(0)}\right\}$; Initial latent image $\boldz^{(0)}$; Trained mappings: $\left\{\boldf_k\right\}$; Scaling factors: $\left\{\hat{s}_k\right\}$; Weights: $\left\{\gamma_k\right\}$, $\alpha$, and $\beta$.}
	\KwOut{Reconstructed images $\left\{\boldxhat_k\right\}$.}  
	\BlankLine
	
	$\boldH_k \gets \left\{\diag\left(\boldA\transp \boldW_k \boldA \bm{1} \right)\right\}, \quad k=1,\dots,K$ ;

	\For{$n = 0$ \KwTo $N_\mathrm{outer}-1$}{
		
		$\boldz^{(n+1)} \gets \argmin_{\boldz}\,  \sum_{k} \frac{1}{2} \gamma_k\left\| \boldf_k(\boldz) - \hat{s}_k \boldx_k^{(n)} \right\|^2_2  +  \alpha H(\boldz)$  \quad \text{\footnotesize (using \acs{LBFGS} initialized with $\boldz^{(n)}$)} ;
		
		\For{$k \gets 1$ \KwTo $K$}{
			$\boldx_k^{(n,0)} \gets \boldx_k^{(n)}$ ;
			
			\For{$m = 0$ \KwTo $N_\mathrm{inner}-1$}{
				$\tilde{\boldx}_k \gets \boldx_k^{(n,m)} - \boldH_k^{-1} \boldA\transp \boldW_k \left(\boldA \boldx_k^{(n,m)}  - \boldb_k\right)$
				$\boldx_k^{(n,m+1)} \gets \left[\frac{\boldH_k \tilde{\boldx}_k  + \beta \gamma_k \hat{s}_k \boldf_k\left(\boldz^{(n+1)}\right)}{\boldH_k + \beta \gamma_k \hat{s}_k^2}\right]_+ $ ;
			}
			$\boldx_k^{(n+1)} \gets \boldx_k^{(n,N_\mathrm{inner})} $ ;
		}
	}
	$\boldxhat_k\gets \boldx_k^{\left(N_\mathrm{outer}\right)}, \quad k=1,\dots,K  $ ;
\end{algorithm}

In addition to image reconstruction, our proposed penalty $R_\mathrm{NN}$ can also be applied for denoising by solving 
\begin{equation}\label{eq:denoise}
	\left\{ \boldx^\star  \right\} = \argmin_{\{\boldx_k\}}  \sum_{k=1}^K \frac{1}{2}\gamma_k\left\| \boldx_k - \boldx^{\mathrm{noisy}}_k  \right\|_2^2  + \beta R_\mathrm{NN}(\left\{ \boldx_k\right\})  ,
\end{equation}
where the weights $\gamma_k$ are used for equal smoothing across channels. The $\boldz$-update is the same as in \eqref{eq:update_z} while the $\boldx_k$-update does not require an iterative algorithm (see Algorithm~\ref{alg:denoise}).

\begin{algorithm}[hbpt]
	\footnotesize
	\caption{Algorithm for multi-energy \ac{CT} denoising with Uconnect}
	\label{alg:denoise}
	\KwIn{Noisy images $\left\{\boldx^\mathrm{noisy}_k\right\}$; Initial latent image $\boldz^{(0)}$; Trained mappings: $\{\boldf_k\}$; Scaling factors: $\left\{\hat{s}_k\right\}$; Weights: $\{\gamma_k\}$, $\alpha$, and $\beta$.}
	\KwOut{Denoised images $\left\{\boldx_k^{\mathrm{denoised}}\right\}$.}  
	\BlankLine
	
	$\boldx_k^{(0)} \gets \boldx^\mathrm{noisy}_k, \quad k=1,\dots,K$ ;

	\For{$n=0$ \KwTo $N-1$}{
		$\boldz^{(n+1)} \gets \argmin_{\boldz}\,  \sum_{k} \frac{1}{2} \gamma_k\left\| \boldf_k(\boldz) - \hat{s}_k \boldx_k^{(n)} \right\|^2_2  +  \alpha H(\boldz)$ \quad \text{\footnotesize (using \acs{LBFGS} initialized with $\boldz^{(n)}$)} ;
		$\boldx_k^{(n+1)} \gets \left[\frac{\boldx_k^\mathrm{noisy} + \beta \boldf_k\left(\boldz^{(n)}\right)}{1 + \beta}\right]_+, \quad k=1,\dots,K$ ;
		
	}
	$\boldx^{\mathrm{denoised}}_k\gets \boldx_k^{(N)}, \quad k=1,\dots,K  $ ;
\end{algorithm}

%% file: unet.tex
\begin{tikzpicture}
\tikzstyle{connection}=[ultra thick,every node/.style={sloped,allow upside down},draw=\edgecolor,opacity=0.7]
\tikzstyle{copyconnection}=[ultra thick,every node/.style={sloped,allow upside down},draw={rgb:blue,4;red,1;green,1;black,3},opacity=0.7]

\node (zero) at (-10,0,0) {};
\node[canvas is zy plane at x=0] (temp) at (-2,0,0) {\includegraphics[width=8cm,height=8cm]{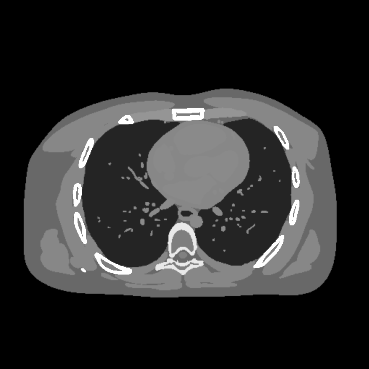}};
\pic[shift={(0,0,0)}] at (0,0,0) {Box={name=cr1,%
		xlabel={{"64",""}},fill=\ConvReluColor,opacity=1,height=40,width=2.5,depth=40}};
\pic[shift={(0,0,0)}] at (cr1-east) {Box={name=cr2,%
		xlabel={{"64",""}},zlabel=512,fill=\ConvReluColor,opacity=1,height=40,width=2.5,depth=40}};
\pic[shift={(1,0,0)}] at (cr2-east) {Box={name=p1,%
		xlabel={{"64",""}},fill=\PoolColor,opacity=1,height=32,width=2.5,depth=32}};
\pic[shift={(0,0,0)}] at (p1-east) {Box={name=cr3,%
		xlabel={{"128",""}},fill=\ConvReluColor,opacity=1,height=32,width=3.5,depth=32}};
\pic[shift={(0,0,0)}] at (cr3-east) {Box={name=cr4,%
		xlabel={{"128",""}},zlabel=256,fill=\ConvReluColor,opacity=1,height=32,width=3.5,depth=32}};
\pic[shift={(0.75,0,0)}] at (cr4-east) {Box={name=p2,%
		xlabel={{"128",""}},fill=\PoolColor,opacity=1,height=25,width=3.5,depth=25}};
\pic[shift={(0,0,0)}] at (p2-east) {Box={name=cr5,%
		xlabel={{"256",""}},fill=\ConvReluColor,opacity=1,height=25,width=4.5,depth=25}};
\pic[shift={(0,0,0)}] at (cr5-east) {Box={name=cr6,%
		xlabel={{"256",""}},zlabel=128,fill=\ConvReluColor,opacity=1,height=25,width=4.5,depth=25}};
\pic[shift={(0.5,0,0)}] at (cr6-east) {Box={name=p3,%
		xlabel={{"256",""}},fill=\PoolColor,opacity=1,height=16,width=4.5,depth=16}};
\pic[shift={(0,0,0)}] at (p3-east) {Box={name=cr7,%
		xlabel={{"512",""}},fill=\ConvReluColor,opacity=1,height=16,width=6,depth=16}};
\pic[shift={(0,0,0)}] at (cr7-east) {Box={name=cr8,%
		xlabel={{"512",""}},zlabel=64,fill=\ConvReluColor,opacity=1,height=16,width=6,depth=16}};
\pic[shift={(0.75,0,0)}] at (cr8-east) {Box={name=p4,%
		xlabel={{"512",""}},fill=\PoolColor,opacity=1,height=8,width=6,depth=8}};
\pic[shift={(0,0,0)}] at (p4-east) {Box={name=cr9,%
		xlabel={{"1024",""}},fill=\ConvReluColor,opacity=1,height=8,width=8,depth=8}};
\pic[shift={(0,0,0)}] at (cr9-east) {Box={name=cr10,%
		xlabel={{"1024",""}},zlabel=32,fill=\ConvReluColor,opacity=1,height=8,width=8,depth=8}};

\pic[shift={(1.2,0,0)}] at (cr10-east) {Box={name=up4,%
		xlabel={{"512",""}},fill=\UnpoolColor,opacity=1,height=16,width=6,depth=16}};
\pic[shift={(0,0,0)}] at (up4-east) {Box={name=cat4,%
		xlabel={{"512",""}},fill=\ConcatColor,opacity=1,height=16,width=6,depth=16}};
\pic[shift={(0,0,0)}] at (cat4-east) {Box={name=ucr8,%
		xlabel={{"512",""}},fill=\ConvReluColor,opacity=1,height=16,width=6,depth=16}};
\pic[shift={(0,0,0)}] at (ucr8-east) {Box={name=ucr7,%
		xlabel={{"512",""}},zlabel=64,fill=\ConvReluColor,opacity=1,height=16,width=6,depth=16}};

\pic[shift={(1.5,0,0)}] at (ucr7-east) {Box={name=up3,%
		xlabel={{"256",""}},fill=\UnpoolColor,opacity=1,height=25,width=4.5,depth=25}};
\pic[shift={(0,0,0)}] at (up3-east) {Box={name=cat3,%
		xlabel={{"256",""}},fill=\ConcatColor,opacity=1,height=25,width=4.5,depth=25}};
\pic[shift={(0,0,0)}] at (cat3-east) {Box={name=ucr6,%
		xlabel={{"256",""}},fill=\ConvReluColor,opacity=1,height=25,width=4.5,depth=25}};
\pic[shift={(0,0,0)}] at (ucr6-east) {Box={name=ucr5,%
		xlabel={{"256",""}},zlabel=128,fill=\ConvReluColor,opacity=1,height=25,width=4.5,depth=25}};

\pic[shift={(1,0,0)}] at (ucr5-east) {Box={name=up2,%
		xlabel={{"128",""}},fill=\UnpoolColor,opacity=1,height=32,width=3.5,depth=32}};
\pic[shift={(0,0,0)}] at (up2-east) {Box={name=cat2,%
		xlabel={{"128",""}},fill=\ConcatColor,opacity=1,height=32,width=3.5,depth=32}};
\pic[shift={(0,0,0)}] at (cat2-east) {Box={name=ucr4,%
		xlabel={{"128",""}},fill=\ConvReluColor,opacity=1,height=32,width=3.5,depth=32}};
\pic[shift={(0,0,0)}] at (ucr4-east) {Box={name=ucr3,%
		xlabel={{"128",""}},zlabel=256,fill=\ConvReluColor,opacity=1,height=32,width=3.5,depth=32}};

\pic[shift={(1.5,0,0)}] at (ucr3-east) {Box={name=up1,%
		xlabel={{"64",""}},fill=\UnpoolColor,opacity=1,height=40,width=2.5,depth=40}};
\pic[shift={(0,0,0)}] at (up1-east) {Box={name=cat1,%
		xlabel={{"64",""}},fill=\ConcatColor,opacity=1,height=40,width=2.5,depth=40}};
\pic[shift={(0,0,0)}] at (cat1-east) {Box={name=ucr2,%
		xlabel={{"64",""}},fill=\ConvReluColor,opacity=1,height=40,width=2.5,depth=40}};
\pic[shift={(0,0,0)}] at (ucr2-east) {Box={name=ucr1,%
		xlabel={{"64",""}},zlabel=512,fill=\ConvReluColor,opacity=1,height=40,width=2.5,depth=40}};

\node[canvas is zy plane at x=36] (temp) at (0,0,0) {\includegraphics[width=8cm,height=8cm]{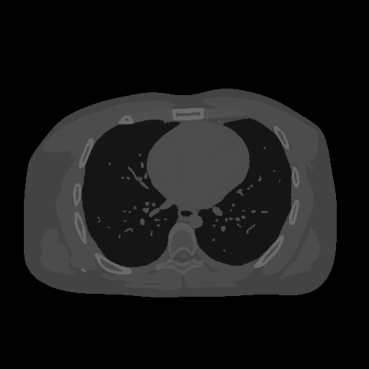}};

\path (cr2-south) -- (cr2-north) coordinate[pos=1.25] (cr2-top) ;
\path (cr4-south) -- (cr4-north) coordinate[pos=1.25] (cr4-top) ;
\path (cr6-south) -- (cr6-north) coordinate[pos=1.25] (cr6-top) ;
\path (cr8-south) -- (cr8-north) coordinate[pos=1.25] (cr8-top) ;

\path (cat4-south)  -- (cat4-north)  coordinate[pos=1.25] (cat4-top) ;
\path (cat3-south)  -- (cat3-north)  coordinate[pos=1.25] (cat3-top) ;
\path (cat2-south)  -- (cat2-north)  coordinate[pos=1.25] (cat2-top)  ;
\path (cat1-south)  -- (cat1-north)  coordinate[pos=1.25] (cat1-top)  ;
\draw [copyconnection]  (cr8-north)  
-- node {\copymidarrow}(cr8-top)
-- node {\copymidarrow}(cat4-top)
-- node {\copymidarrow} (cat4-north);
\draw [copyconnection]  (cr6-north)  
-- node {\copymidarrow}(cr6-top)
-- node {\copymidarrow}(cat3-top)
-- node {\copymidarrow} (cat3-north);
\draw [copyconnection]  (cr4-north)  
-- node {\copymidarrow}(cr4-top)
-- node {\copymidarrow}(cat2-top)
-- node {\copymidarrow} (cat2-north);
\draw [copyconnection]  (cr2-north)  
-- node {\copymidarrow}(cr2-top)
-- node {\copymidarrow}(cat1-top)
-- node {\copymidarrow} (cat1-north);

\coordinate (O_legend) at (13.7,-3);
\pic[shift={(0,0,0)}] at (11.5,-4) {Box={name=cr11,caption=conv 3$\times$3 + BN + ReLU,%
		fill=\ConvReluColor,opacity=1,height=3,width=3,depth=3}};
\pic[shift={(0,0,0)}] at (11.5,-5) {Box={name=cr11,caption=max pool 2$\times$2,%
		fill=\PoolColor,opacity=1,height=3,width=3,depth=3}};
\pic[shift={(0,0,0)}] at (11.5,-6) {Box={name=cr11,caption=up-conv 2$\times$2,%
		fill=\UnpoolColor,opacity=1,height=3,width=3,depth=3}};
\pic[shift={(0,0,0)}] at (11.5,-7) {Box={name=cr11,caption=concatenation,%
		fill=\ConcatColor,opacity=1,height=3,width=3,depth=3}};

\end{tikzpicture}

%% file: results.tex
\section{Experiments}\label{sec:experiments}

We conducted experiments on two types of data: synthetic data and real patient images. In both cases, we considered \ac{2D} \ac{CT} images at $K=6$ energies:  $E_1 = 40$~keV, $E_2 = 60$~keV, $E_3 = 80$~keV, $E_4 = 100$~keV, $E_5 = 120$~keV and  $E_6 = 140$~keV. 

U-Net architectures were built in Python 3.8 with Tensorflow 2.10 using NVIDIA GeForce GTX 1660 Ti (6 GB memory). In both experiments, the models were trained on one dataset, and were tested on a separate dataset of the same type. 

We compared the approaches mentioned in Section~\ref{sec:method}, \ie{} Huber, \ac{JTV}, \ac{DTV}, \ac{ASSIST}, as well as our proposed Uconnect method with $H = R_\mathrm{H}$ (Huber), for two different tasks: image reconstruction in Section~\ref{sec:XCATexperiment} and image denoising in Section~\ref{sec:patient}. In addition, we also conducted a comparison with a \ac{DL} approach called \ac{RED-CNN} \cite{chen2017low} in terms of image denoising task for patient data in Section~\ref{sec:patient}. We used a \ac{SQS} algorithm \cite{elbakri2002statistical} for the individual reconstructions  with Huber penalty and the Chambolle-Pock algorithm \cite{chambolle2011first,sidky2012convex} for \ac{JTV} and \ac{DTV}. All algorithms were implemented in Python programming language.

The objective measures   \ac{SSIM} and \ac{PSNR}, with respect to a reference image $\boldx^{\mathrm{ref}} = \left[ x_1^{\mathrm{ref}},\dots,x_J^{\mathrm{ref}} \right]\transp$,  were employed for quantitative evaluation of the image quality. We utilized the functions \texttt{structural\_similarity} and \texttt{peak\_signal\_noise\_ratio} from the Python package \texttt{skimage.metrics}.

\subsection{Synthetic data}\label{sec:XCATexperiment}

\subsubsection{Data Preparation, Training, and Setup}

We used the \ac{XCAT} software \cite{segars2008realistic} to generate 2 phantoms $512\times{}512\times{}280$, one male and one female, at the different energies described above. We used the $L = 280$ slices of the male phantom to train the $\boldf_k$s throughout 700 epochs. The batch size was set to 5 and the learning rate was set to $2\cdot 10^{-4}$. We considered the female phantom as the \ac{GT} $\left\{ \boldx_k^{\mathrm{gt}}  \right\}$ for reconstruction experiments.

For each energy, we utilized a constant intensity $\bar{h}_k=5000$  to simulate low-dose (low photon count) measurements $\{\boldy_k\}$ following \eqref{eq:poisson2}  (with $\boldx_k=\boldx_k^{\mathrm{gt}}$) with $N_\rms=120$ projection angles and $N_\rmd=750$ detectors using a fan beam projector within the ASTRA toolbox \cite{vanAarle:16}. 

The prior image used for \ac{DTV} and \ac{ASSIST} was reconstructed using the unbinned data $\boldy = [y_1,\dots,y_I]\transp$, $y_i = \sum_k y_{i,k}$ and the Huber penalty, i.e., as
\begin{equation}\label{eq:xprior}
    \boldx^\mathrm{prior} = \argmin_{\boldx \in \R^J_+}\, \frac{1}{2}\left\| \boldA \boldx - \boldb \right\|^2_{\boldW} + \beta R_\mathrm{H}(\boldx)
\end{equation}
with $\boldb = [b_i,\dots,b_I]\transp$, $b_i= \log \left(\sum_{k=1}^K \bar{h}_k / y_i\right)$, $\boldW = \mathrm{diag}\left\{\boldy\right\}$ and $\beta = 300$,  using a \ac{SQS}-based iterative algorithm. For \ac{ASSIST}, we searched 48 most similar overlapped patches of size $12\times{}12$ with a stride of 6 in a search window of $10\times{}10$ to construct tensor units. For \ac{DTV}, we set $\eta=0.7$ and $\epsilon=10^{-5}$ in \eqref{eq:dtv_xi}. For Uconnect, we manually set $\alpha=1$ in \eqref{eq:update_z}, $N_\mathrm{outer}=50$ and $N_\mathrm{inner}=50$ in Algorithm~\ref{alg:recon}.

\subsubsection{Pre-evaluation}\label{sec:pre-eval-xcat}

\begin{figure}[!htb]
    \centering
	\subfigure[$\boldzhat_1$]{
		\includegraphics[width=0.45\linewidth]{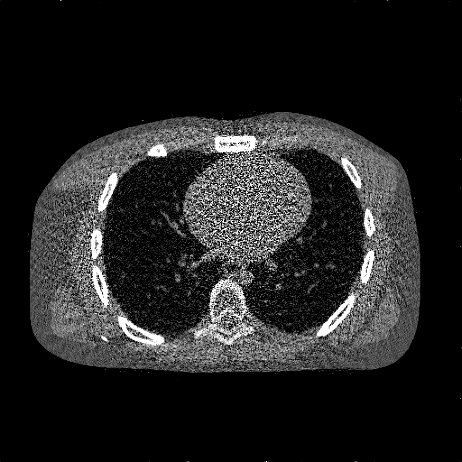}
	}%
    \subfigure[$\boldzhat_3$]{
		\includegraphics[width=0.45\linewidth]{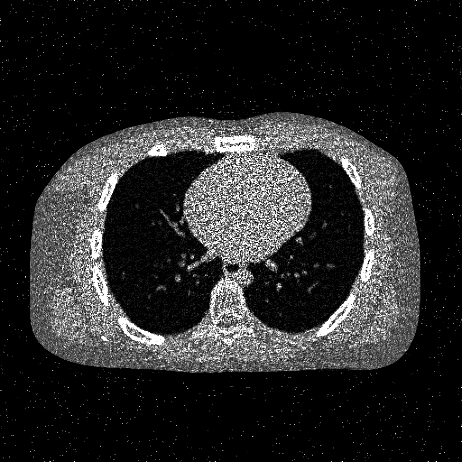}
	}%

    \subfigure[$\boldzhat_6$]{
		\includegraphics[width=0.45\linewidth]{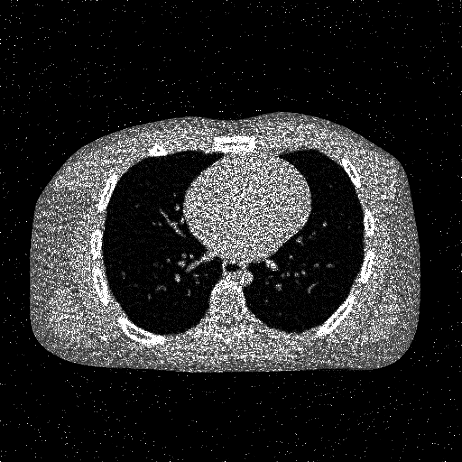}
	}%
    \subfigure[$\boldzhat_{\mathrm{all}}$]{
		\includegraphics[width=0.45\linewidth]{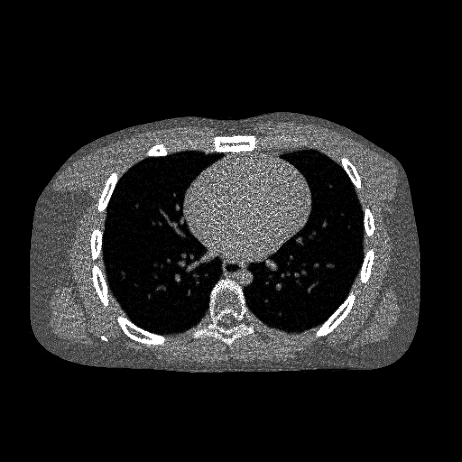}
	}%
    \caption{Latent variables $\boldzhat_1$, $\boldzhat_3$ and $\boldzhat_6$ obtained from \eqref{eq:z_1_k}, and $\boldzhat_\mathrm{all}$ obtained from \eqref{eq:z_init} (using the scout reconstructions $\boldxhat_k$).  }\label{fig:xcat512zhat}
\end{figure}

\begin{figure}[!htb]
	\centering
	\subfigure[$\boldf_1\left(\boldzhat_1\right)$]{
		\begin{overpic}[width=0.3\linewidth]{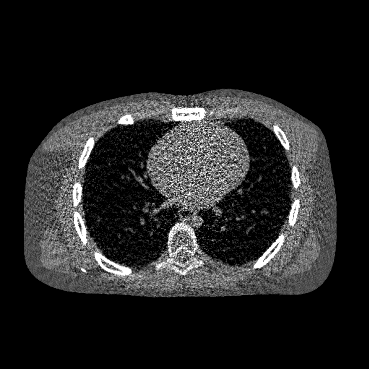}
			\put(1,90){\footnotesize \textbf{\color{red}{\ac{SSIM}=0.9925}}}
			\put(1,1){\footnotesize \textbf{\color{red}{\ac{PSNR}=21.210 dB}}}
		\end{overpic}
	}%
	\subfigure[$\boldf_3\left(\boldzhat_3\right)$]{
		\begin{overpic}[width=0.3\linewidth]{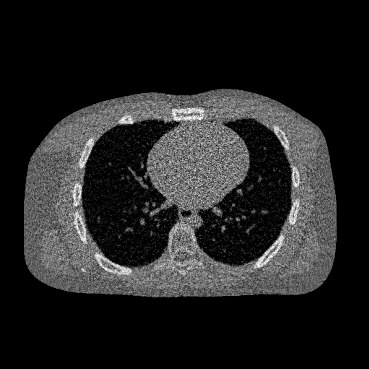}
			\put(1,90){\footnotesize \textbf{\color{red}{\ac{SSIM}=0.9983}}}
			\put(1,1){\footnotesize \textbf{\color{red}{\ac{PSNR}=19.855 dB}}}
		\end{overpic}
	}%
	\subfigure[$\boldf_6\left(\boldzhat_6\right)$]{
		\begin{overpic}[width=0.3\linewidth]{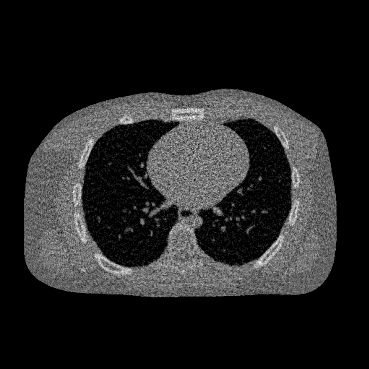}
			\put(1,90){\footnotesize \textbf{\color{red}{\ac{SSIM}=0.9990}}}
			\put(1,1){\footnotesize \textbf{\color{red}{\ac{PSNR}=19.569 dB}}}
		\end{overpic}
	}%
	
	\subfigure[$\boldf_1\left(\boldzhat_\mathrm{all}\right)$]{
		\begin{overpic}[width=0.3\linewidth]{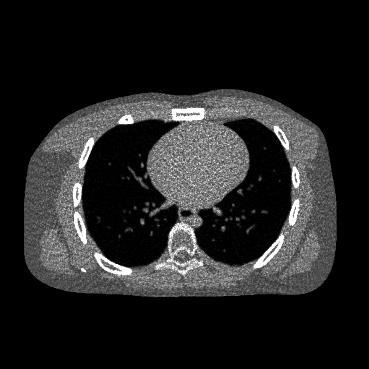}
			\put(1,90){\footnotesize \textbf{\color{red}{\ac{SSIM}=0.9969}}}
			\put(1,1){\footnotesize \textbf{\color{red}{\ac{PSNR}=25.134 dB}}}
		\end{overpic}
	}%
	\subfigure[$\boldf_3\left(\boldzhat_\mathrm{all}\right)$]{
		\begin{overpic}[width=0.3\linewidth]{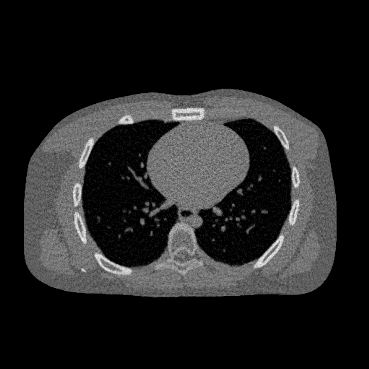}
			\put(1,90){\footnotesize \textbf{\color{red}{\ac{SSIM}=0.9995}}}
			\put(1,1){\footnotesize \textbf{\color{red}{\ac{PSNR}=25.365 dB}}}
		\end{overpic}
	}%
	\subfigure[$\boldf_6\left(\boldzhat_\mathrm{all}\right)$]{
		\begin{overpic}[width=0.3\linewidth]{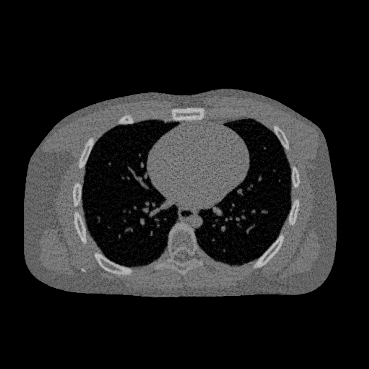}
			\put(1,90){\footnotesize \textbf{\color{red}{\ac{SSIM}=0.9997}}}
			\put(1,1){\footnotesize \textbf{\color{red}{\ac{PSNR}=25.250 dB}}}
		\end{overpic}
	}%
	\caption{Generated images $\boldf_1(\boldzhat_1)$, $\boldf_3(\boldzhat_3)$ and $\boldf_6(\boldzhat_6)$ obtained from \eqref{eq:z_1_k}, and $\boldf_k(\boldzhat_\mathrm{all})$ obtained from \eqref{eq:z_init} (using the scout reconstructions $\boldxhat_k$).  }\label{fig:xcat512z}
\end{figure}

Uconnect is motivated by the hypothesis that combining the data from all energy bins may help in reducing noise. To verify this, we investigated the quality of the generated image $\boldf_k(\boldz)$ used in $R_\mathrm{NN}$ \eqref{eq:r-final}, where the latent variable $\boldz$ is obtained using all energy bins, compared with the image obtained from each bin $k$ individually, without the regularizer $H$ (\ie{} $\alpha=0$), that is to say
\begin{equation}\label{eq:z_init}
	\boldzhat_{\mathrm{all}} = \argmin_{\boldz\in\R^J_+}\,  \sum_{k=1}^K \frac{1}{2} \gamma_k\left\| \boldf_k(\boldz) - \hat{s}_k \boldxhat_k \right\|^2_2 
\end{equation}
and 
\begin{equation}\label{eq:z_1_k}
	\boldzhat_k = \argmin_{\boldz\in\R^J_+}\, \frac{1}{2}\left\| \boldf_k(\boldz) - \hat{s}_k \boldxhat_k \right\|^2_2 
\end{equation}  
where the images $\boldxhat_k$ are scout reconstructions obtained by \ac{WLS} (i.e., solving \eqref{eq:pwls} with $\beta=0$) with \ac{SQS} and 1000 iterations. Note that solving \eqref{eq:z_1_k} corresponds to ``inverting'' $\boldf_k$. Fig.\@~\ref{fig:xcat512zhat} shows the estimated latent variables $\boldzhat_{\mathrm{all}}$ and $\boldzhat_k$, $k=1,3,6$, and we observe that using all channels reduces the noise. The generated images $\boldf_k(\boldzhat_{\mathrm{all}})$ and $\boldf_k(\boldzhat_k )$ are shown alongside their \ac{SSIM} and \ac{PSNR} (using $\boldx^\mathrm{gt}_k$ as the reference images) in Fig.\@~\ref{fig:xcat512z}. We observe that both \ac{SSIM} and \ac{PSNR} values of $\boldf_k(\boldzhat_{\mathrm{all}} )$ is higher than those of each $\boldf_k(\boldzhat_k )$, which supports our initial hypothesis.   


\subsubsection{Reconstruction Results}\label{sec:img}

\begin{figure}[!htb]
	\settoheight{\tempdima}{\includegraphics[height=0.28\linewidth]{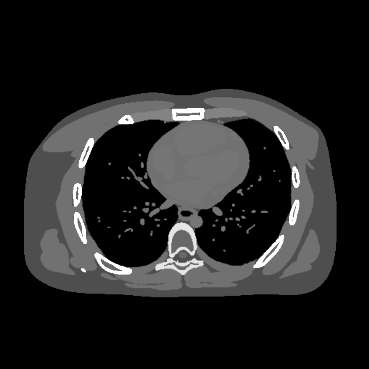}}%
	\centering
	
	\begin{tabular}{@{}c@{}c@{\hspace{-0.0cm}}c@{\hspace{-0.0cm}}c@{\hspace{-0.0cm}}c@{\hspace{-0.0cm}}c@{}}
		& \scriptsize $k=1$ (40 keV) & \scriptsize $k=3$ (80 keV) & \scriptsize $k=6$ (140 keV)  \vspace{-0.1cm} \\
		
		\rowname{\ac{GT}} & 
		\begin{tikzpicture}
		\begin{scope}[spy using outlines={rectangle,yellow,magnification=2,size=7mm,connect spies}]
		\node {\includegraphics[height=\tempdima]{./xcat_GT_40keV}};
		\spy on (-0.85,-0.45) in node [left] at (-0.4,1.0);
		\spy on (0.02,0.45) in node [left] at (0.4,1.0);
		\spy on (-0.03,-0.43) in node [right] at (0.5,1.0);
		\end{scope}
		\end{tikzpicture}
		&
		\begin{tikzpicture}
		\begin{scope}[spy using outlines={rectangle,yellow,magnification=2,size=7mm,connect spies}]
		\node {\includegraphics[height=\tempdima]{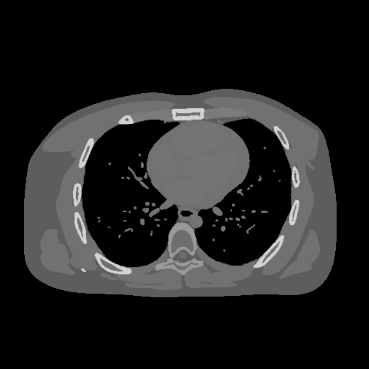}};
		\spy on (-0.85,-0.45) in node [left] at (-0.4,1.0);
		\spy on (0.02,0.45) in node [left] at (0.4,1.0);
		\spy on (-0.03,-0.43) in node [right] at (0.5,1.0);  
		\end{scope}
		\end{tikzpicture} 
		&
		\begin{tikzpicture}
		\begin{scope}[spy using outlines={rectangle,yellow,magnification=2,size=7mm,connect spies}]
		\node {\includegraphics[height=\tempdima]{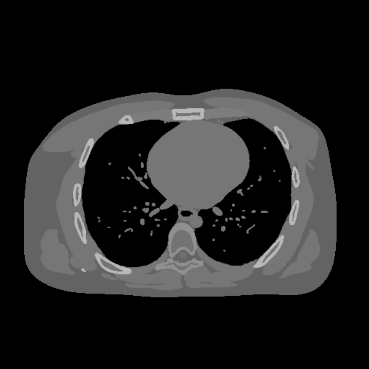}};
		\spy on (-0.85,-0.45) in node [left] at (-0.4,1.0);
		\spy on (0.02,0.45) in node [left] at (0.4,1.0);
		\spy on (-0.03,-0.43) in node [right] at (0.5,1.0);   
		\end{scope}
		\end{tikzpicture} 
		\\
		
		\rowname{Huber} & 
		\begin{tikzpicture}
		\begin{scope}[spy using outlines={rectangle,yellow,magnification=2,size=7mm,connect spies}]
		\node {\includegraphics[height=\tempdima]{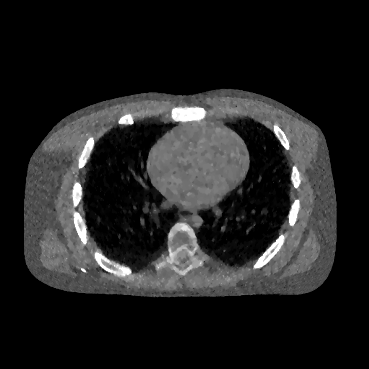}};
		\spy on (-0.85,-0.45) in node [left] at (-0.4,1.0);
		\spy on (0.02,0.45) in node [left] at (0.4,1.0);
		\spy on (-0.03,-0.43) in node [right] at (0.5,1.0);
		\end{scope}
		\end{tikzpicture}
		&
		\begin{tikzpicture}
		\begin{scope}[spy using outlines={rectangle,yellow,magnification=2,size=7mm,connect spies}]
		\node {\includegraphics[height=\tempdima]{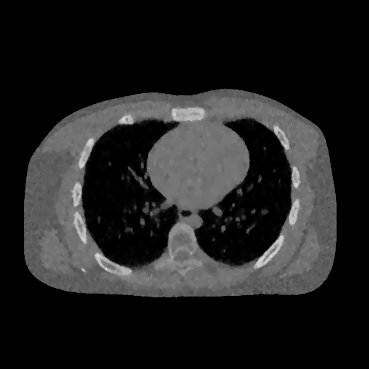}};
		\spy on (-0.85,-0.45) in node [left] at (-0.4,1.0);
		\spy on (0.02,0.45) in node [left] at (0.4,1.0);
		\spy on (-0.03,-0.43) in node [right] at (0.5,1.0);   
		\end{scope}
		\end{tikzpicture} 
		&
		\begin{tikzpicture}
		\begin{scope}[spy using outlines={rectangle,yellow,magnification=2,size=7mm,connect spies}]
		\node {\includegraphics[height=\tempdima]{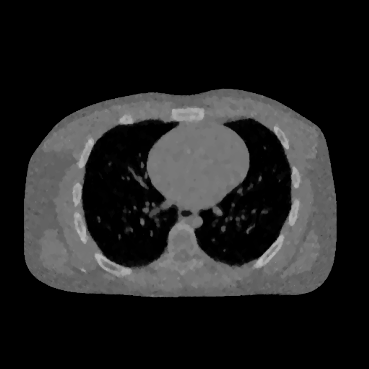}};
		\spy on (-0.85,-0.45) in node [left] at (-0.4,1.0);
		\spy on (0.02,0.45) in node [left] at (0.4,1.0);
		\spy on (-0.03,-0.43) in node [right] at (0.5,1.0);
		\end{scope}
		\end{tikzpicture} 
		\\
		
		\rowname{\ac{DTV}} & 
		\begin{tikzpicture}
		\begin{scope}[spy using outlines={rectangle,yellow,magnification=2,size=7mm,connect spies}]
		\node {\includegraphics[height=\tempdima]{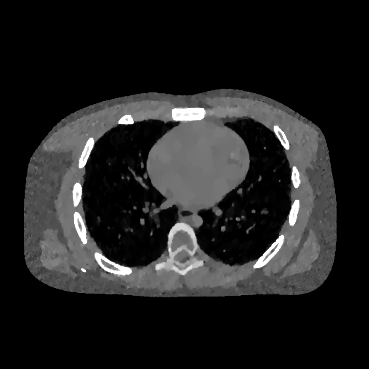}};
		\spy on (-0.85,-0.45) in node [left] at (-0.4,1.0);
		\spy on (0.02,0.45) in node [left] at (0.4,1.0);
		\spy on (-0.03,-0.43) in node [right] at (0.5,1.0);
		\end{scope}
		\end{tikzpicture}
		&
		\begin{tikzpicture}
		\begin{scope}[spy using outlines={rectangle,yellow,magnification=2,size=7mm,connect spies}]
		\node {\includegraphics[height=\tempdima]{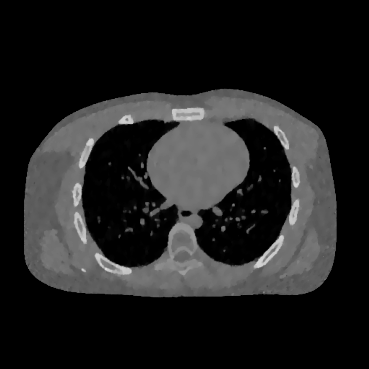}};
		\spy on (-0.85,-0.45) in node [left] at (-0.4,1.0);
		\spy on (0.02,0.45) in node [left] at (0.4,1.0);
		\spy on (-0.03,-0.43) in node [right] at (0.5,1.0); 
		\end{scope}
		\end{tikzpicture} 
		&
		\begin{tikzpicture}
		\begin{scope}[spy using outlines={rectangle,yellow,magnification=2,size=7mm,connect spies}]
		\node {\includegraphics[height=\tempdima]{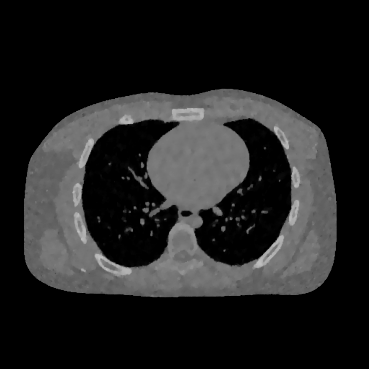}};
		\spy on (-0.85,-0.45) in node [left] at (-0.4,1.0);
		\spy on (0.02,0.45) in node [left] at (0.4,1.0);
		\spy on (-0.03,-0.43) in node [right] at (0.5,1.0);  
		\end{scope}
		\end{tikzpicture} 
		\\
		
		\rowname{\ac{JTV}} & 
		\begin{tikzpicture}
		\begin{scope}[spy using outlines={rectangle,yellow,magnification=2,size=7mm,connect spies}]
		\node {\includegraphics[height=\tempdima]{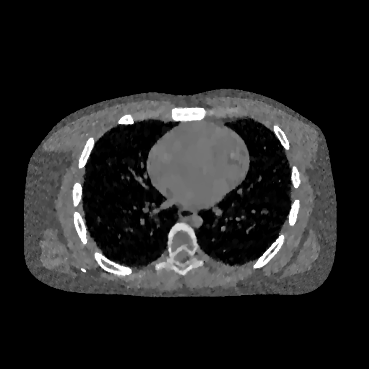}};
		\spy on (-0.85,-0.45) in node [left] at (-0.4,1.0);
		\spy on (0.02,0.45) in node [left] at (0.4,1.0);
		\spy on (-0.03,-0.43) in node [right] at (0.5,1.0); 
		\end{scope}
		\end{tikzpicture}
		&
		\begin{tikzpicture}
		\begin{scope}[spy using outlines={rectangle,yellow,magnification=2,size=7mm,connect spies}]
		\node {\includegraphics[height=\tempdima]{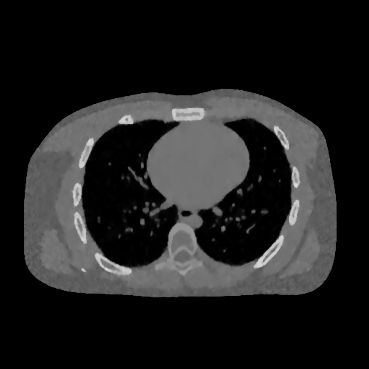}};
		\spy on (-0.85,-0.45) in node [left] at (-0.4,1.0);
		\spy on (0.02,0.45) in node [left] at (0.4,1.0);
		\spy on (-0.03,-0.43) in node [right] at (0.5,1.0);   
		\end{scope}
		\end{tikzpicture} 
		&
		\begin{tikzpicture}
		\begin{scope}[spy using outlines={rectangle,yellow,magnification=2,size=7mm,connect spies}]
		\node {\includegraphics[height=\tempdima]{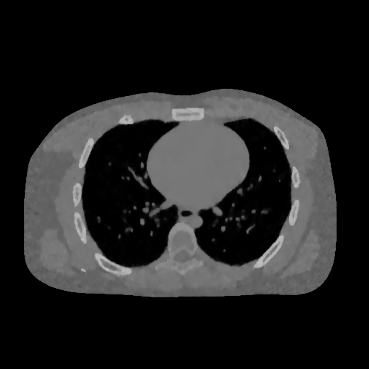}};
		\spy on (-0.85,-0.45) in node [left] at (-0.4,1.0);
		\spy on (0.02,0.45) in node [left] at (0.4,1.0);
		\spy on (-0.03,-0.43) in node [right] at (0.5,1.0);  
		\end{scope}
		\end{tikzpicture} 
		\\
		
		\rowname{\ac{ASSIST}} & 
		\begin{tikzpicture}
		\begin{scope}[spy using outlines={rectangle,yellow,magnification=2,size=7mm,connect spies}]
		\node {\includegraphics[height=\tempdima]{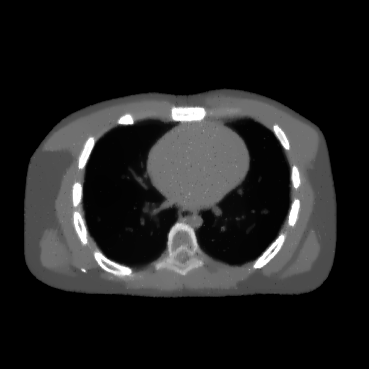}};
		\spy on (-0.85,-0.45) in node [left] at (-0.4,1.0);
		\spy on (0.02,0.45) in node [left] at (0.4,1.0);
		\spy on (-0.03,-0.43) in node [right] at (0.5,1.0); 
		\end{scope}
		\end{tikzpicture}
		&
		\begin{tikzpicture}
		\begin{scope}[spy using outlines={rectangle,yellow,magnification=2,size=7mm,connect spies}]
		\node {\includegraphics[height=\tempdima]{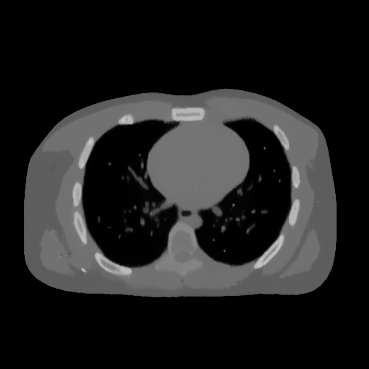}};
		\spy on (-0.85,-0.45) in node [left] at (-0.4,1.0);
		\spy on (0.02,0.45) in node [left] at (0.4,1.0);
		\spy on (-0.03,-0.43) in node [right] at (0.5,1.0);   
		\end{scope}
		\end{tikzpicture} 
		&
		\begin{tikzpicture}
		\begin{scope}[spy using outlines={rectangle,yellow,magnification=2,size=7mm,connect spies}]
		\node {\includegraphics[height=\tempdima]{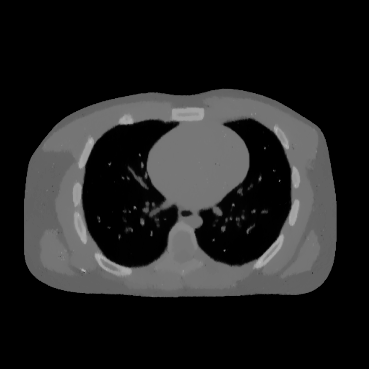}};
		\spy on (-0.85,-0.45) in node [left] at (-0.4,1.0);
		\spy on (0.02,0.45) in node [left] at (0.4,1.0);
		\spy on (-0.03,-0.43) in node [right] at (0.5,1.0); 
		\end{scope}
		\end{tikzpicture} 
		\\
		
		\rowname{Uconnect} & 
		\begin{tikzpicture}
		\begin{scope}[spy using outlines={rectangle,yellow,magnification=2,size=7mm,connect spies}]
		\node {\includegraphics[height=\tempdima]{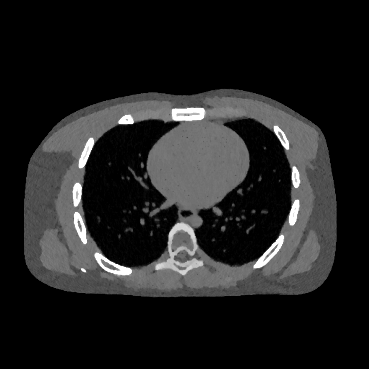}};
		\spy on (-0.85,-0.45) in node [left] at (-0.4,1.0);
		\spy on (0.02,0.45) in node [left] at (0.4,1.0);
		\spy on (-0.03,-0.43) in node [right] at (0.5,1.0);
		\end{scope}
		\end{tikzpicture}
		&
		\begin{tikzpicture}
		\begin{scope}[spy using outlines={rectangle,yellow,magnification=2,size=7mm,connect spies}]
		\node {\includegraphics[height=\tempdima]{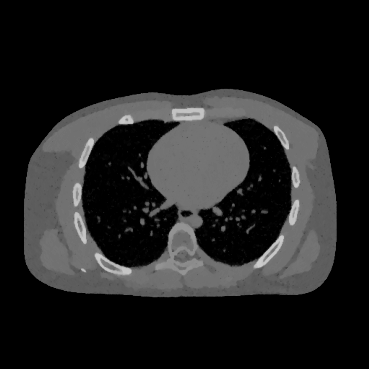}};
		\spy on (-0.85,-0.45) in node [left] at (-0.4,1.0);
		\spy on (0.02,0.45) in node [left] at (0.4,1.0);
		\spy on (-0.03,-0.43) in node [right] at (0.5,1.0); 
		\end{scope}
		\end{tikzpicture} 
		&
		\begin{tikzpicture}
		\begin{scope}[spy using outlines={rectangle,yellow,magnification=2,size=7mm,connect spies}]
		\node {\includegraphics[height=\tempdima]{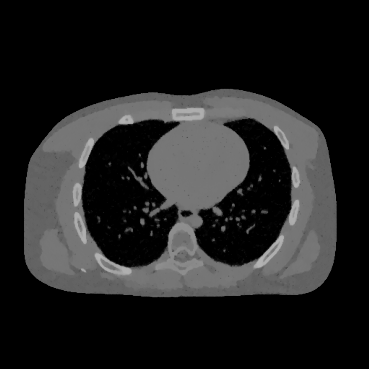}};
		\spy on (-0.85,-0.45) in node [left] at (-0.4,1.0);
		\spy on (0.02,0.45) in node [left] at (0.4,1.0);
		\spy on (-0.03,-0.43) in node [right] at (0.5,1.0);   
		\end{scope}
		\end{tikzpicture} 
		\vspace{-1.9cm}
		\\
		\rowname{} & 
		\begin{tikzpicture}
		\node {\includegraphics[width=\tempdima]{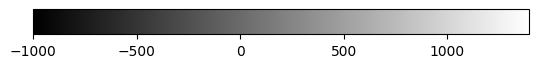}};
		\end{tikzpicture}
		&
		\begin{tikzpicture}
		\node {\includegraphics[width=\tempdima]{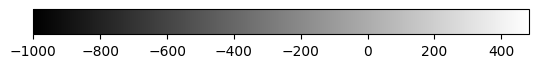}};
		\end{tikzpicture} 
		&
		\begin{tikzpicture}
		\node {\includegraphics[width=\tempdima]{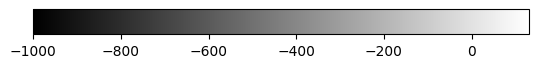}};
		\end{tikzpicture}

	\end{tabular}	
	\caption{\Ac{GT} and reconstructed images at $E_1=40$, $E_2=80$  and $E_3=140$~keV using different methods.}\label{fig:xcatrecon}
\end{figure}

The images reconstructed by different approaches for the \ac{XCAT} phantom at 40, 80 and 140~keV (respectively from left to right columns) with zoomed-in areas marked by yellow boxes are shown in Fig.\@~\ref{fig:xcatrecon} in \ac{HU} value. Their respective chosen optimal parameter $\beta$ corresponds to the highest \ac{PSNR} in Fig.~\ref{fig:psnr_xcat} below. It is visually evident that the results of Huber, \ac{DTV} and \ac{JTV} remain noisy and tend to over-smooth bone structures. \ac{ASSIST} tends to oversmooth the entire image especially at 140~keV, despite good denoising performance. In contrast, our proposed method Uconnect is capable of both preserving features and reducing noise.

We also evaluated quantitatively the aforementioned methods via the index \ac{PSNR} and \ac{SSIM}, using the \ac{GT} images as the reference. Images were reconstructed with different values of the regularization $\beta$, from $\beta=\beta_\mathrm{min}$ to $\beta_\mathrm{max}$, where $\beta_\mathrm{min}$ and $\beta_\mathrm{max}$ are different for each method due to the variation in order of magnitude of the different penalties.

\begin{figure}[!htb]
	\centering
	\ref{named2} \\
	\subfigure[40~keV]{
		\begin{minipage}[t]{0.5\linewidth}
			\begin{tikzpicture}
			\begin{axis}[
			height=0.7\linewidth,
			width=1.1\linewidth,
			tick label style={font=\scriptsize},
			x label style={at={(axis description cs:0.5,0.2)},font=\scriptsize},
			y label style={at={(axis description cs:0.13,0.5)},anchor=south,font=\scriptsize},
			xlabel={$\beta/\beta_\mathrm{max}$},
			ylabel={SSIM},
			scaled y ticks=false,
			y tick label style={/pgf/number format/.cd, fixed, precision=4}, 
			legend columns=-1,
			legend entries={Uconnect,\ac{ASSIST},\ac{JTV},\ac{DTV},Huber},
			legend to name=named2,
			]
			\addplot[color=blue, mark=otimes*, mark size=1.5pt] table[x=beta/max_beta, y=SSIM, col sep=comma] {./xcat_ssim_Uconnect_40keV.txt};
			\addplot[color=violet, mark=diamond*, mark size=1.5pt] table[x=beta/max_beta, y=SSIM, col sep=comma] {./xcat_ssim_ASSIST_40keV.txt};
			\addplot[color=red, mark=diamond*, mark size=1.5pt] table[x=beta/max_beta, y=SSIM, col sep=comma] {./xcat_ssim_JTV_40keV.txt};
			\addplot[color=green, mark=diamond*, mark size=1.5pt] table[x=beta/max_beta, y=SSIM, col sep=comma] {./xcat_ssim_dTV_40keV.txt};
			\addplot[color=orange, mark=diamond*, mark size=1.5pt] table[x=beta/max_beta, y=SSIM, col sep=comma] {./xcat_ssim_Huber_40keV.txt};
			\end{axis}
			\end{tikzpicture}
		\end{minipage}%
	}%
	\subfigure[60~keV]{
		\begin{minipage}[t]{0.5\linewidth}
			\begin{tikzpicture}
			\begin{axis}[
			height=0.7\linewidth,
			width=1.1\linewidth,
			tick label style={font=\scriptsize},
			x label style={at={(axis description cs:0.5,0.2)},font=\scriptsize},
			xlabel={$\beta/\beta_\mathrm{max}$},
			scaled y ticks=false,
			y tick label style={/pgf/number format/.cd, fixed, precision=4},
			]
			\addplot[color=blue, mark=otimes*, mark size=1.5pt] table[x=beta/max_beta, y=SSIM, col sep=comma] {./xcat_ssim_Uconnect_60keV.txt};
			\addplot[color=violet, mark=diamond*, mark size=1.5pt] table[x=beta/max_beta, y=SSIM, col sep=comma] {./xcat_ssim_ASSIST_60keV.txt};
			\addplot[color=red, mark=diamond*, mark size=1.5pt] table[x=beta/max_beta, y=SSIM, col sep=comma] {./xcat_ssim_JTV_60keV.txt};
			\addplot[color=green, mark=diamond*, mark size=1.5pt] table[x=beta/max_beta, y=SSIM, col sep=comma] {./xcat_ssim_dTV_60keV.txt};
			\addplot[color=orange, mark=diamond*, mark size=1.5pt] table[x=beta/max_beta, y=SSIM, col sep=comma] {./xcat_ssim_Huber_60keV.txt};
			\end{axis}
			\end{tikzpicture}
		\end{minipage}%
	}
	\subfigure[80~keV]{
		\begin{minipage}[t]{0.5\linewidth}
			\begin{tikzpicture}
			\begin{axis}[
			height=0.7\linewidth,
			width=1.1\linewidth,
			tick label style={font=\scriptsize},
			x label style={at={(axis description cs:0.5,0.2)},font=\scriptsize},
			y label style={at={(axis description cs:0.12,0.5)},anchor=south,font=\scriptsize},
			xlabel={$\beta/\beta_\mathrm{max}$},
			ylabel={SSIM},
			scaled y ticks=false,
			y tick label style={/pgf/number format/.cd, fixed, precision=4},
			]
			\addplot[color=blue, mark=otimes*, mark size=1.5pt] table[x=beta/max_beta, y=SSIM, col sep=comma] {./xcat_ssim_Uconnect_80keV.txt};
			\addplot[color=violet, mark=diamond*, mark size=1.5pt] table[x=beta/max_beta, y=SSIM, col sep=comma] {./xcat_ssim_ASSIST_80keV.txt};
			\addplot[color=red, mark=diamond*, mark size=1.5pt] table[x=beta/max_beta, y=SSIM, col sep=comma] {./xcat_ssim_JTV_80keV.txt};
			\addplot[color=green, mark=diamond*, mark size=1.5pt] table[x=beta/max_beta, y=SSIM, col sep=comma] {./xcat_ssim_dTV_80keV.txt};
			\addplot[color=orange, mark=diamond*, mark size=1.5pt] table[x=beta/max_beta, y=SSIM, col sep=comma] {./xcat_ssim_Huber_80keV.txt};
			\end{axis}
			\end{tikzpicture}
		\end{minipage}%
	}%
	\subfigure[100~keV]{
		\begin{minipage}[t]{0.5\linewidth}
			\begin{tikzpicture}
			\begin{axis}[
			height=0.7\linewidth,
			width=1.1\linewidth,
			tick label style={font=\scriptsize},
			x label style={at={(axis description cs:0.5,0.2)},font=\scriptsize},
			y label style={at={(axis description cs:0.2,0.5)},anchor=south,font=\scriptsize},
			xlabel={$\beta/\beta_\mathrm{max}$},
			scaled y ticks=false,
			y tick label style={/pgf/number format/.cd, fixed, precision=4},
			]
			\addplot[color=blue, mark=otimes*, mark size=1.5pt] table[x=beta/max_beta, y=SSIM, col sep=comma] {./xcat_ssim_Uconnect_100keV.txt};
			\addplot[color=violet, mark=diamond*, mark size=1.5pt] table[x=beta/max_beta, y=SSIM, col sep=comma] {./xcat_ssim_ASSIST_100keV.txt};
			\addplot[color=red, mark=diamond*, mark size=1.5pt] table[x=beta/max_beta, y=SSIM, col sep=comma] {./xcat_ssim_JTV_100keV.txt};
			\addplot[color=green, mark=diamond*, mark size=1.5pt] table[x=beta/max_beta, y=SSIM, col sep=comma] {./xcat_ssim_dTV_100keV.txt};
			\addplot[color=orange, mark=diamond*, mark size=1.5pt] table[x=beta/max_beta, y=SSIM, col sep=comma] {./xcat_ssim_Huber_100keV.txt};
			\end{axis}
			\end{tikzpicture}
		\end{minipage}%
	}
	\subfigure[120~keV]{
		\begin{minipage}[t]{0.5\linewidth}
			\begin{tikzpicture}
			\begin{axis}[
			height=0.7\linewidth,
			width=1.1\linewidth,
			tick label style={font=\scriptsize},
			x label style={at={(axis description cs:0.5,0.2)},font=\scriptsize},
			y label style={at={(axis description cs:0.12,0.5)},anchor=south,font=\scriptsize},
			xlabel={$\beta/\beta_\mathrm{max}$},
			ylabel={SSIM},
			scaled y ticks=false,
			y tick label style={/pgf/number format/.cd, fixed, precision=4},
			]
			\addplot[color=blue, mark=otimes*, mark size=1.5pt] table[x=beta/max_beta, y=SSIM, col sep=comma] {./xcat_ssim_Uconnect_120keV.txt};
			\addplot[color=violet, mark=diamond*, mark size=1.5pt] table[x=beta/max_beta, y=SSIM, col sep=comma] {./xcat_ssim_ASSIST_120keV.txt};
			\addplot[color=red, mark=diamond*, mark size=1.5pt] table[x=beta/max_beta, y=SSIM, col sep=comma] {./xcat_ssim_JTV_120keV.txt};
			\addplot[color=green, mark=diamond*, mark size=1.5pt] table[x=beta/max_beta, y=SSIM, col sep=comma] {./xcat_ssim_dTV_120keV.txt};
			\addplot[color=orange, mark=diamond*, mark size=1.5pt] table[x=beta/max_beta, y=SSIM, col sep=comma] {./xcat_ssim_Huber_120keV.txt};
			\end{axis}
			\end{tikzpicture}
		\end{minipage}%
	}%
	\subfigure[140~keV]{
		\begin{minipage}[t]{0.5\linewidth}
			\begin{tikzpicture}
			\begin{axis}[
			height=0.7\linewidth,
			width=1.1\linewidth,
			tick label style={font=\scriptsize},
			x label style={at={(axis description cs:0.5,0.2)},font=\scriptsize},
			y label style={at={(axis description cs:0.12,0.5)},anchor=south,font=\scriptsize},
			xlabel={$\beta/\beta_\mathrm{max}$},
			scaled y ticks=false,
			y tick label style={/pgf/number format/.cd, fixed, precision=4},
			]
			\addplot[color=blue, mark=otimes*, mark size=1.5pt] table[x=beta/max_beta, y=SSIM, col sep=comma] {./xcat_ssim_Uconnect_140keV.txt};
			\addplot[color=violet, mark=diamond*, mark size=1.5pt] table[x=beta/max_beta, y=SSIM, col sep=comma] {./xcat_ssim_ASSIST_140keV.txt};
			\addplot[color=red, mark=diamond*, mark size=1.5pt] table[x=beta/max_beta, y=SSIM, col sep=comma] {./xcat_ssim_JTV_140keV.txt};
			\addplot[color=green, mark=diamond*, mark size=1.5pt] table[x=beta/max_beta, y=SSIM, col sep=comma] {./xcat_ssim_dTV_140keV.txt};
			\addplot[color=orange, mark=diamond*, mark size=1.5pt] table[x=beta/max_beta, y=SSIM, col sep=comma] {./xcat_ssim_Huber_140keV.txt};
			\end{axis}
			\end{tikzpicture}
		\end{minipage}%
	}%
	\caption{\Ac{SSIM} values of the reconstructed images using different methods.}\label{fig:ssim_xcat}
\end{figure}
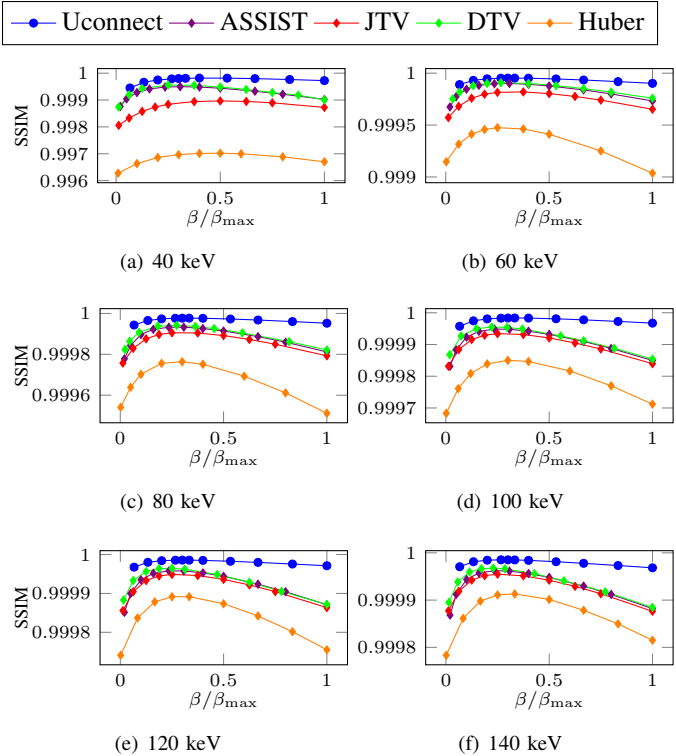

Fig.\@~\ref{fig:psnr_xcat} shows the \ac{PSNR} values of the reconstructed images at each of the 6 energy bins using different approaches. The images using Uconnect show the highest \ac{PSNR} values at all energies, with gains between 2.5 and 5.5 dB in \ac{PSNR} compared to the best performance by the other approaches at each energy. Similar results in terms of \ac{SSIM} are observed (Fig.\@~\ref{fig:ssim_xcat}). These quantitative results are consistent with the above visual comparison.

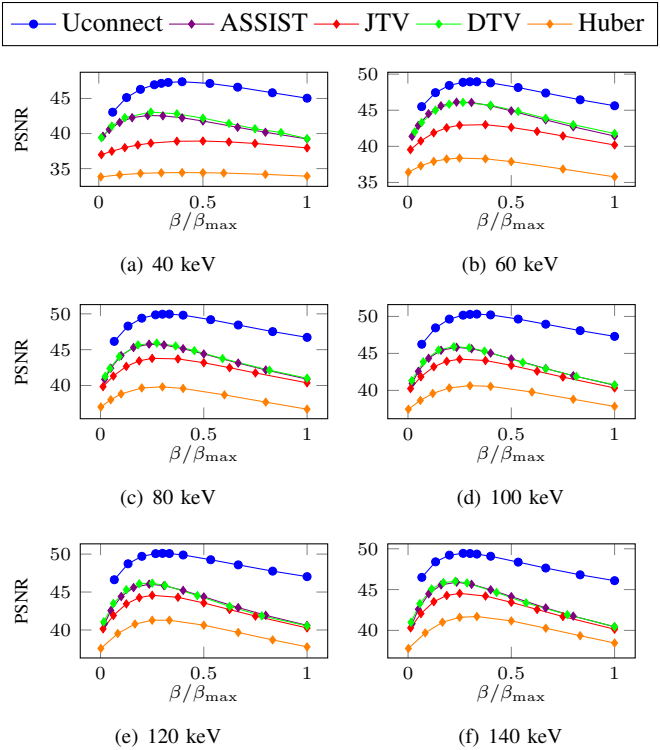
\begin{figure}[!hpb]
	\centering
	\ref{named2} \\
	\subfigure[40~keV]{
		\begin{minipage}[t]{0.5\linewidth}
			\begin{tikzpicture}
			\begin{axis}[
			height=0.7\linewidth,
			width=1.1\linewidth,
			tick label style={font=\scriptsize},
			x label style={at={(axis description cs:0.5,0.2)},font=\scriptsize},
			y label style={at={(axis description cs:0.2,0.5)},anchor=south,font=\scriptsize},
			xlabel={$\beta/\beta_\mathrm{max}$},
			ylabel={PSNR},
			legend columns=-1,
			legend entries={Uconnect,\ac{ASSIST},\ac{JTV},\ac{DTV},Huber},
			legend to name=named2,
			]
			\addplot[color=blue, mark=otimes*, mark size=1.5pt] table[x=beta/max_beta, y=PSNR, col sep=comma] {./xcat_psnr_Uconnect_40keV.txt};
			\addplot[color=violet, mark=diamond*, mark size=1.5pt] table[x=beta/max_beta, y=PSNR, col sep=comma] {./xcat_psnr_ASSIST_40keV.txt};
			\addplot[color=red, mark=diamond*, mark size=1.5pt] table[x=beta/max_beta, y=PSNR, col sep=comma] {./xcat_psnr_JTV_40keV.txt};
			\addplot[color=green, mark=diamond*, mark size=1.5pt] table[x=beta/max_beta, y=PSNR, col sep=comma] {./xcat_psnr_dTV_40keV.txt};
			\addplot[color=orange, mark=diamond*, mark size=1.5pt] table[x=beta/max_beta, y=PSNR, col sep=comma] {./xcat_psnr_Huber_40keV.txt};
			\end{axis}
			\end{tikzpicture}
		\end{minipage}%
	}%
	\subfigure[60~keV]{
		\begin{minipage}[t]{0.5\linewidth}
			\begin{tikzpicture}
			\begin{axis}[
			height=0.7\linewidth,
			width=1.1\linewidth,
			tick label style={font=\scriptsize},
			x label style={at={(axis description cs:0.5,0.2)},font=\scriptsize},
			xlabel={$\beta/\beta_\mathrm{max}$},
			]
			\addplot[color=blue, mark=otimes*, mark size=1.5pt] table[x=beta/max_beta, y=PSNR, col sep=comma] {./xcat_psnr_Uconnect_60keV.txt};
			\addplot[color=violet, mark=diamond*, mark size=1.5pt] table[x=beta/max_beta, y=PSNR, col sep=comma] {./xcat_psnr_ASSIST_60keV.txt};
			\addplot[color=red, mark=diamond*, mark size=1.5pt] table[x=beta/max_beta, y=PSNR, col sep=comma] {./xcat_psnr_JTV_60keV.txt};
			\addplot[color=green, mark=diamond*, mark size=1.5pt] table[x=beta/max_beta, y=PSNR, col sep=comma] {./xcat_psnr_dTV_60keV.txt};
			\addplot[color=orange, mark=diamond*, mark size=1.5pt] table[x=beta/max_beta, y=PSNR, col sep=comma] {./xcat_psnr_Huber_60keV.txt};
			\end{axis}
			\end{tikzpicture}
		\end{minipage}%
	}
	\subfigure[80~keV]{
		\begin{minipage}[t]{0.5\linewidth}
			\begin{tikzpicture}
			\begin{axis}[
			height=0.7\linewidth,
			width=1.1\linewidth,
			tick label style={font=\scriptsize},
			x label style={at={(axis description cs:0.5,0.2)},font=\scriptsize},
			y label style={at={(axis description cs:0.2,0.5)},anchor=south,font=\scriptsize},
			xlabel={$\beta/\beta_\mathrm{max}$},
			ylabel={PSNR},
			]
			\addplot[color=blue, mark=otimes*, mark size=1.5pt] table[x=beta/max_beta, y=PSNR, col sep=comma] {./xcat_psnr_Uconnect_80keV.txt};
			\addplot[color=violet, mark=diamond*, mark size=1.5pt] table[x=beta/max_beta, y=PSNR, col sep=comma] {./xcat_psnr_ASSIST_80keV.txt};
			\addplot[color=red, mark=diamond*, mark size=1.5pt] table[x=beta/max_beta, y=PSNR, col sep=comma] {./xcat_psnr_JTV_80keV.txt};
			\addplot[color=green, mark=diamond*, mark size=1.5pt] table[x=beta/max_beta, y=PSNR, col sep=comma] {./xcat_psnr_dTV_80keV.txt};
			\addplot[color=orange, mark=diamond*, mark size=1.5pt] table[x=beta/max_beta, y=PSNR, col sep=comma] {./xcat_psnr_Huber_80keV.txt};
			\end{axis}
			\end{tikzpicture}
		\end{minipage}%
	}%
	\subfigure[100~keV]{
		\begin{minipage}[t]{0.5\linewidth}
			\begin{tikzpicture}
			\begin{axis}[
			height=0.7\linewidth,
			width=1.1\linewidth,
			tick label style={font=\scriptsize},
			x label style={at={(axis description cs:0.5,0.2)},font=\scriptsize},
			y label style={at={(axis description cs:0.2,0.5)},anchor=south,font=\scriptsize},
			xlabel={$\beta/\beta_\mathrm{max}$},
			]
			\addplot[color=blue, mark=otimes*, mark size=1.5pt] table[x=beta/max_beta, y=PSNR, col sep=comma] {./xcat_psnr_Uconnect_100keV.txt};
			\addplot[color=violet, mark=diamond*, mark size=1.5pt] table[x=beta/max_beta, y=PSNR, col sep=comma] {./xcat_psnr_ASSIST_100keV.txt};
			\addplot[color=red, mark=diamond*, mark size=1.5pt] table[x=beta/max_beta, y=PSNR, col sep=comma] {./xcat_psnr_JTV_100keV.txt};
			\addplot[color=green, mark=diamond*, mark size=1.5pt] table[x=beta/max_beta, y=PSNR, col sep=comma] {./xcat_psnr_dTV_100keV.txt};
			\addplot[color=orange, mark=diamond*, mark size=1.5pt] table[x=beta/max_beta, y=PSNR, col sep=comma] {./xcat_psnr_Huber_100keV.txt};
			\end{axis}
			\end{tikzpicture}
		\end{minipage}%
	}
	\subfigure[120~keV]{
		\begin{minipage}[t]{0.5\linewidth}
			\begin{tikzpicture}
			\begin{axis}[
			height=0.7\linewidth,
			width=1.1\linewidth,
			tick label style={font=\scriptsize},
			x label style={at={(axis description cs:0.5,0.2)},font=\scriptsize},
			y label style={at={(axis description cs:0.2,0.5)},anchor=south,font=\scriptsize},
			xlabel={$\beta/\beta_\mathrm{max}$},
			ylabel={PSNR},
			]
			\addplot[color=blue, mark=otimes*, mark size=1.5pt] table[x=beta/max_beta, y=PSNR, col sep=comma] {./xcat_psnr_Uconnect_120keV.txt};
			\addplot[color=violet, mark=diamond*, mark size=1.5pt] table[x=beta/max_beta, y=PSNR, col sep=comma] {./xcat_psnr_ASSIST_120keV.txt};
			\addplot[color=red, mark=diamond*, mark size=1.5pt] table[x=beta/max_beta, y=PSNR, col sep=comma] {./xcat_psnr_JTV_120keV.txt};
			\addplot[color=green, mark=diamond*, mark size=1.5pt] table[x=beta/max_beta, y=PSNR, col sep=comma] {./xcat_psnr_dTV_120keV.txt};
			\addplot[color=orange, mark=diamond*, mark size=1.5pt] table[x=beta/max_beta, y=PSNR, col sep=comma] {./xcat_psnr_Huber_120keV.txt};
			\end{axis}
			\end{tikzpicture}
		\end{minipage}%
	}%
	\subfigure[140~keV]{
		\begin{minipage}[t]{0.5\linewidth}
			\begin{tikzpicture}
			\begin{axis}[
			height=0.7\linewidth,
			width=1.1\linewidth,
			tick label style={font=\scriptsize},
			x label style={at={(axis description cs:0.5,0.2)},font=\scriptsize},
			y label style={at={(axis description cs:0.2,0.5)},anchor=south,font=\scriptsize},
			xlabel={$\beta/\beta_\mathrm{max}$},
			]
			\addplot[color=blue, mark=otimes*, mark size=1.5pt] table[x=beta/max_beta, y=PSNR, col sep=comma] {./xcat_psnr_Uconnect_140keV.txt};
			\addplot[color=violet, mark=diamond*, mark size=1.5pt] table[x=beta/max_beta, y=PSNR, col sep=comma] {./xcat_psnr_ASSIST_140keV.txt};
			\addplot[color=red, mark=diamond*, mark size=1.5pt] table[x=beta/max_beta, y=PSNR, col sep=comma] {./xcat_psnr_JTV_140keV.txt};
			\addplot[color=green, mark=diamond*, mark size=1.5pt] table[x=beta/max_beta, y=PSNR, col sep=comma] {./xcat_psnr_dTV_140keV.txt};
			\addplot[color=orange, mark=diamond*, mark size=1.5pt] table[x=beta/max_beta, y=PSNR, col sep=comma] {./xcat_psnr_Huber_140keV.txt};
			\end{axis}
			\end{tikzpicture}
		\end{minipage}%
	}%
	\caption{\Ac{PSNR} values of the reconstructed images using different methods.}\label{fig:psnr_xcat}
\end{figure}

\subsection{Patient Data}\label{sec:patient}

\subsubsection{Data Preparation, Training, and Setup}
2 collections of real patient data images of size $512 \times 512$ from 2 different patients were obtained by the Philips IQon Spectral \ac{CT} scanner from the Poitiers University Hospital, Poitiers, France. The Philips IQon Spectral \ac{CT} is a dual energy scanner that produces 2 images at 80~kVp and 120~kVp from which 6 \acp{VMI} were obtained by interpolation (same energies as in Section~\ref{sec:XCATexperiment}). $L = 360$ slices of the first dataset were used to train the $\boldf_k$s throughout 500 epochs. The batch size was set to 5 and the learning rate was set to 0.0002.

Owing to the fact that these images do not represent \acp{GT} attenuation of real patients, reconstructing from simulated measurements $\{\boldy_k\}$ with Poisson noise added,  as we did with the \ac{XCAT} phantom, was not considered. Therefore we assessed the different penalties for a denoising task with a least-squares data-fidelity term, i.e. 
\begin{equation}
	L_k(\boldx_k) = \frac{1}{2} \left\| \boldx_k - \boldx_k^\mathrm{noisy}\right\|^2_2
\end{equation}
where $\boldx_k^\mathrm{noisy}$ is the noisy image simulated as
\begin{equation}\label{eq:noise}
	\boldx_k^\mathrm{noisy} = \boldx_k^\mathrm{patient} + \bm{\varepsilon}_k,
\quad \bm{\varepsilon}_k \sim \mathcal{N}\left(\bm{0},\,\bm{\Sigma}_k^{2}\right),
\end{equation} 
and $\boldx_k^\mathrm{patient}$ a patient image from the second dataset, and  $\bm{\Sigma}_k^{2} = \diag\left\{\sigma_{1,k}^2,\dots,\sigma_{J,k}^2\right\}$. The Uconnect approach for denoising corresponds to solving \eqref{eq:denoise}. The other methods were implemented by using the same approach with different penalties.   

We estimated realistic $\sigma_{j,k}$ values with a Monte-Carlo approach. We created $Q=30$ independent measurements with random Poisson noise following \eqref{eq:poisson2} with $\boldx_k = \left\{\boldx_k^\mathrm{patient}\right\}$ for each energy $k=1,\dots,K$, with intensity $\bar{h}_{k} = \frac{2}{3} \cdot 10^5$ for all $k$, from which we reconstructed the spectral images $\boldx_k^{[q]}=\left[x_{1,k}^{[q]},\dots,x_{J,k}^{[q]}\right]\transp$, $k=1,\dots,K$ and $q=1,\dots,Q$, by \ac{WLS} (i.e., \eqref{eq:pwls} with $\beta=0$).  The variance $\sigma_{j,k}^2$ at each pixel $j$ for each energy $k$ was estimated as
\begin{equation}\label{eq:variance}
	\sigma_{j,k}^2 = \frac{1}{Q-1} \sum_{q=1}^{Q} \left(x_{j,k}^{[q]} - \xbar_{j,k}\right)^2, \quad \xbar_{j,k} = \frac{1}{Q} \sum_{q=1}^{Q}  x_{j,k}^{[q]} 
\end{equation}

The weights $\gamma_k$ in our method (i.e., Eq.~\eqref{eq:r-final} and \eqref{eq:denoise}) were defined as
\begin{equation}\label{eq:weight_denoise}
	\gamma_k = \frac{\frac{1}{\sum_{j=1}^{J} \sigma_{j,k}^2}}{\sum_{k=1}^{K} \frac{1}{\sum_{j=1}^{J} \sigma_{j,k}^2}}
\end{equation}
in order to give more weight to the less noisy images. Although these weights are in principle unknown, they can be estimated \cite{fessler1996mean}.

For the \ac{DL}-based denoising method \ac{RED-CNN}, a total of $K$ networks need to be trained for each energy $k=1,\dots,K$. The training dataset for each energy $k$ consists of a collection of $L=360$ pairs $\left\{\left(\boldx^\mathrm{noisy}_{k,l}, \boldx^\mathrm{patient}_{k,l}\right)\right\}_{l=1}^{L}$, where the noisy input image $\boldx^\mathrm{noisy}_{k,l}$ is simulated as
\begin{equation}\label{eq:noiseREDCNN}
    \boldx^\mathrm{noisy}_{k,l} = \boldx^\mathrm{patient}_{k,l} + \bm{\varepsilon}_{k,l},
    \quad \bm{\varepsilon}_{k,l} \sim \mathcal{N}\left(\bm{0},\,\bm{\Sigma}_{k,l}^{2}\right),
\end{equation} 
where the $l$\th{} patient image from the first dataset $\boldx^\mathrm{patient}_{k,l}$ serves as the corresponding target image during training, and $\bm{\Sigma}_{k,l}^{2} = \diag\left\{\sigma_{1,k,l}^2,\dots,\sigma_{J,k,l}^2\right\}$. The estimation of $\sigma_{j,k,l}^2$ follows the same procedure as described in \eqref{eq:variance} for each $l$, employing $Q=30$ independent measurements in a Monte-Carlo approach. The patch size in  patch-based training of \ac{RED-CNN}  was set to $55\times{}55$, with an initial base learning rate set at $10^{-5}$, gradually decreasing to $10^{-8}$ over the course of 10000 epochs.

The prior image $\boldx^\mathrm{prior}$ for \ac{DTV} and \ac{ASSIST} was defined as the weighted average across energy bins, i.e.,
\begin{equation}\label{eq:variance}
	\boldx^\mathrm{prior} = \frac{1}{K} \sum_{k=1}^K  \gamma_k  \boldx_k^\mathrm{noisy}   \, .
\end{equation}
In \ac{ASSIST}, tensor units were constructed by searching 12 most similar patches of size $24\times{}24$ with a stride of 12 in the search window of $20\times{}20$. We used the same parameter setting as with the simulated XCAT phantom datasets, $\eta=0.7$ in \eqref{eq:dtv_xi} for \ac{DTV}, and for Uconnect, $\alpha$ in \eqref{eq:update_z} was set manually to 0.4 and $N$ in Algorithm~\ref{alg:denoise} was set to 50.

\subsubsection{Pre-evaluation}\label{sec:pre-eval-real}

We proceeded with the same evaluation as in Section~\ref{sec:pre-eval-xcat}, but using  $\boldx_k^\mathrm{noisy}$ instead of $\boldxhat_k$ in \eqref{eq:z_init} and \eqref{eq:z_1_k}, and using $\boldx_k^\mathrm{patient}$ as the reference for \ac{SSIM} and \ac{PSNR}. 

\begin{figure}[!htb]
	\centering
	\subfigure[$\boldf_1\left(\boldzhat_1\right)$]{
		\begin{overpic}[width=0.3\linewidth]{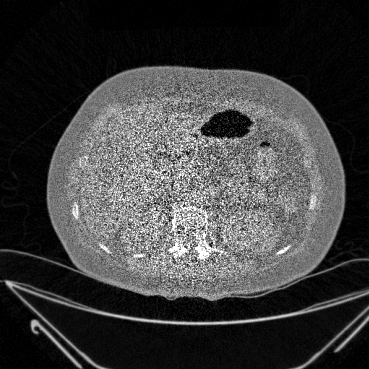}
			\put(1,90){\footnotesize \textbf{\color{red}{\ac{SSIM}=0.9951}}}
			\put(1,1){\footnotesize \textbf{\color{red}{\ac{PSNR}=22.443 dB}}}
		\end{overpic}
	}%
	\subfigure[$\boldf_3\left(\boldzhat_3\right)$]{
		\begin{overpic}[width=0.3\linewidth]{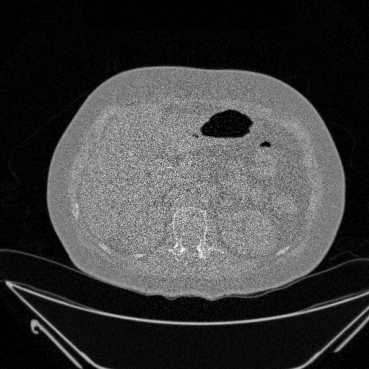}
			\put(1,90){\footnotesize \textbf{\color{red}{\ac{SSIM}=0.9994}}}
			\put(1,1){\footnotesize \textbf{\color{red}{\ac{PSNR}=24.750 dB}}}
		\end{overpic}
	}%
	\subfigure[$\boldf_6\left(\boldzhat_6\right)$]{
		\begin{overpic}[width=0.3\linewidth]{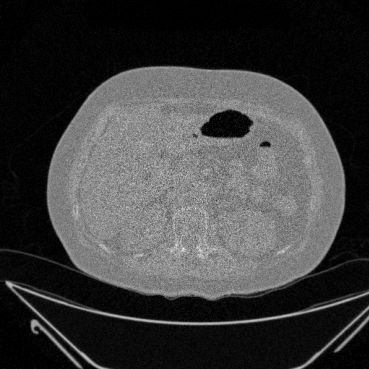}
			\put(1,90){\footnotesize \textbf{\color{red}{\ac{SSIM}=0.9997}}}
			\put(1,1){\footnotesize \textbf{\color{red}{\ac{PSNR}=25.826 dB}}}
		\end{overpic}
	}%
	
	\subfigure[$\boldf_1\left(\boldzhat_{\mathrm{all}}\right)$]{
		\begin{overpic}[width=0.3\linewidth]{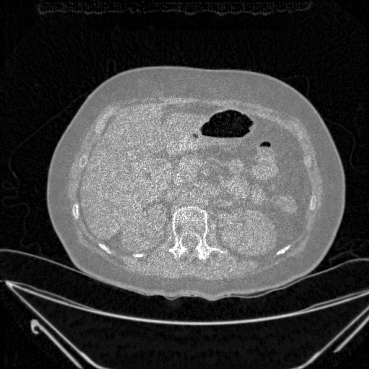}
			\put(1,90){\footnotesize \textbf{\color{red}{\ac{SSIM}=0.9983}}}
			\put(1,1){\footnotesize \textbf{\color{red}{\ac{PSNR}=29.081 dB}}}
		\end{overpic}
	}%
	\subfigure[$\boldf_3\left(\boldzhat_{\mathrm{all}}\right)$]{
		\begin{overpic}[width=0.3\linewidth]{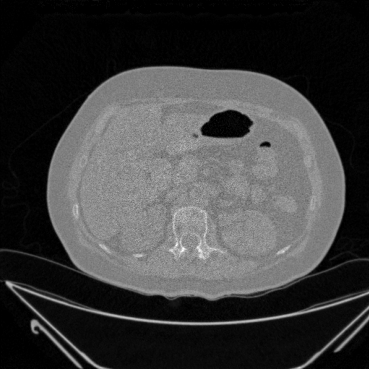}
			\put(1,90){\footnotesize \textbf{\color{red}{\ac{SSIM}=0.9999}}}
			\put(1,1){\footnotesize \textbf{\color{red}{\ac{PSNR}=31.668 dB}}}
		\end{overpic}
	}%
	\subfigure[$\boldf_6\left(\boldzhat_{\mathrm{all}}\right)$]{
		\begin{overpic}[width=0.3\linewidth]{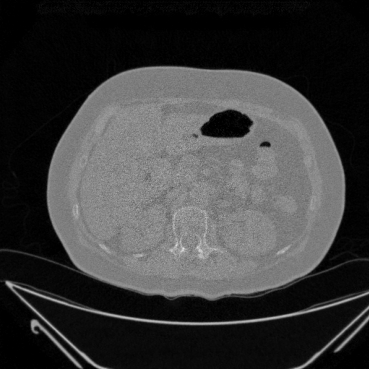}
			\put(1,90){\footnotesize \textbf{\color{red}{\ac{SSIM}=0.9999}}}
			\put(1,1){\footnotesize \textbf{\color{red}{\ac{PSNR}=31.291 dB}}}
		\end{overpic}
	}%
	\caption{$\boldf_1(\boldzhat_1)$, $\boldf_3(\boldzhat_3)$ and $\boldf_6(\boldzhat_6)$ obtained from \eqref{eq:z_1_k}, and $\boldf_k(\boldzhat_\mathrm{all})$ obtained from \eqref{eq:z_init} (using the noisy images $\boldx^\mathrm{noisy}_k$).  }\label{fig:realdata512z}
\end{figure}

Fig.\@~\ref{fig:realdata512z} displays the obtained images $\boldf_k(\boldzhat_{\mathrm{all}} )$ and $\boldf_k(\boldzhat_k )$, $k=1,3,6$ with their corresponding \ac{SSIM} and \ac{PSNR}. Similarly to the results of the simulated images, each $\boldf_k(\boldzhat_{\mathrm{all}} )$ is less noisy and has higher \ac{SSIM} and \ac{PSNR} compared to the corresponding $\boldf_k(\boldzhat_k )$.

\subsubsection{Denoising Results}

\begin{figure}[htbp]
	\settoheight{\tempdima}{\includegraphics[height=0.280\linewidth]{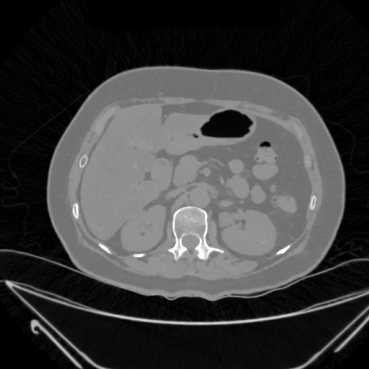}}%
	\centering
	
	\begin{tabular}{@{}c@{}c@{\hspace{-0.0cm}}c@{\hspace{-0.0cm}}c@{\hspace{-0.0cm}}c@{\hspace{-0.0cm}}c@{}}
		& \scriptsize $k=1$ (40 keV) & \scriptsize $k=3$ (80 keV) & \scriptsize $k=6$ (140 keV)  \vspace{-0.1cm} \\

		\rowname{$\boldx_k^\mathrm{patient}$} & 
		\begin{tikzpicture}
		\begin{scope}[spy using outlines={rectangle,yellow,magnification=2,size=7mm,connect spies}]
		\node {\includegraphics[height=\tempdima]{./realdata_GT_40keV}};
		\spy on (-0.05,0.1) in node [left] at (-0.35,1.1);
		\spy on (0.05,-0.305) in node [right] at (0.35,1.1);   
		\end{scope}
		\end{tikzpicture}
		&
		\begin{tikzpicture}
		\begin{scope}[spy using outlines={rectangle,yellow,magnification=2,size=7mm,connect spies}]
		\node {\includegraphics[height=\tempdima]{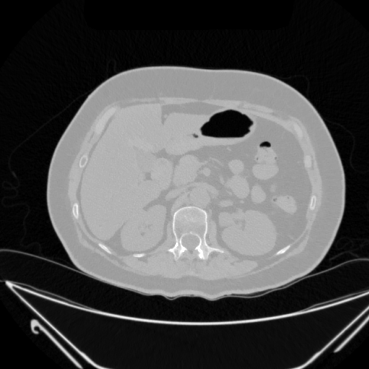}};
		\spy on (-0.05,0.1) in node [left] at (-0.35,1.1);
		\spy on (0.05,-0.305) in node [right] at (0.35,1.1);     
		\end{scope}
		\end{tikzpicture} 
		&
		\begin{tikzpicture}
		\begin{scope}[spy using outlines={rectangle,yellow,magnification=2,size=7mm,connect spies}]
		\node {\includegraphics[height=\tempdima]{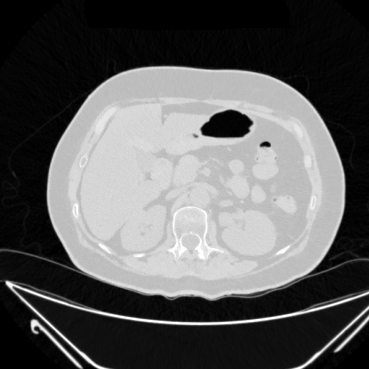}};
		\spy on (-0.05,0.1) in node [left] at (-0.35,1.1);
		\spy on (0.05,-0.305) in node [right] at (0.35,1.1);     
		\end{scope}
		\end{tikzpicture} 
		\vspace{-0.16cm}\\
		
		\rowname{$\boldx_k^\mathrm{noisy}$} & 
		\begin{tikzpicture}
		\begin{scope}[spy using outlines={rectangle,yellow,magnification=2,size=7mm,connect spies}]
		\node {\includegraphics[height=\tempdima]{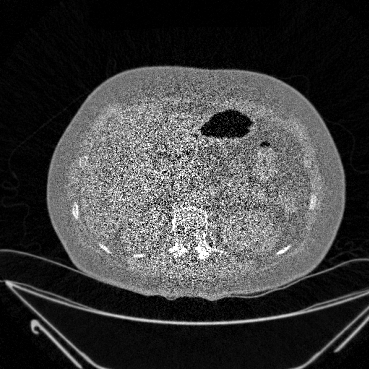}};
		\spy on (-0.05,0.1) in node [left] at (-0.35,1.1);
		\spy on (0.05,-0.305) in node [right] at (0.35,1.1);   
		\end{scope}
		\end{tikzpicture}
		&
		\begin{tikzpicture}
		\begin{scope}[spy using outlines={rectangle,yellow,magnification=2,size=7mm,connect spies}]
		\node {\includegraphics[height=\tempdima]{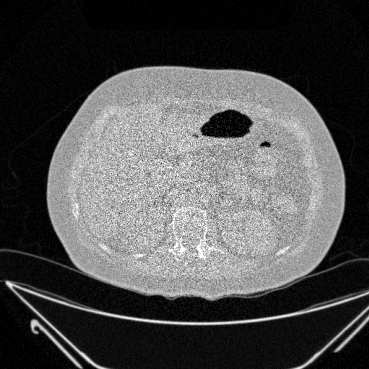}};
		\spy on (-0.05,0.1) in node [left] at (-0.35,1.1);
		\spy on (0.05,-0.305) in node [right] at (0.35,1.1);     
		\end{scope}
		\end{tikzpicture} 
		&
		\begin{tikzpicture}
		\begin{scope}[spy using outlines={rectangle,yellow,magnification=2,size=7mm,connect spies}]
		\node {\includegraphics[height=\tempdima]{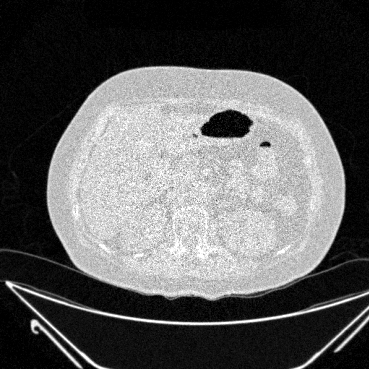}};
		\spy on (-0.05,0.1) in node [left] at (-0.35,1.1);
		\spy on (0.05,-0.305) in node [right] at (0.35,1.1);     
		\end{scope}
		\end{tikzpicture} 
		\vspace{-0.16cm}\\		
		
		\rowname{Huber} & 
		\begin{tikzpicture}
		\begin{scope}[spy using outlines={rectangle,yellow,magnification=2,size=7mm,connect spies}]
		\node {\includegraphics[height=\tempdima]{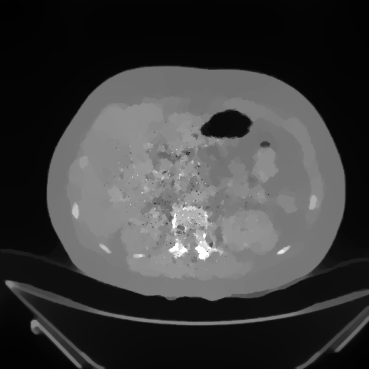}};
		\spy on (-0.05,0.1) in node [left] at (-0.35,1.1);
		\spy on (0.05,-0.305) in node [right] at (0.35,1.1);  
		\end{scope}
		\end{tikzpicture}
		&
		\begin{tikzpicture}
		\begin{scope}[spy using outlines={rectangle,yellow,magnification=2,size=7mm,connect spies}]
		\node {\includegraphics[height=\tempdima]{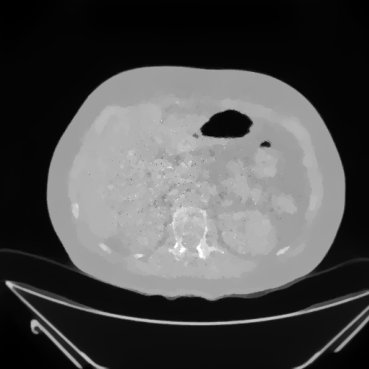}};
		\spy on (-0.05,0.1) in node [left] at (-0.35,1.1);
		\spy on (0.05,-0.305) in node [right] at (0.35,1.1);     
		\end{scope}
		\end{tikzpicture} 
		&
		\begin{tikzpicture}
		\begin{scope}[spy using outlines={rectangle,yellow,magnification=2,size=7mm,connect spies}]
		\node {\includegraphics[height=\tempdima]{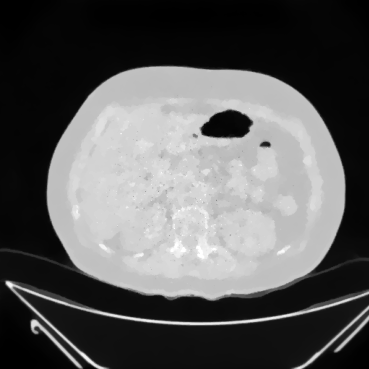}};
		\spy on (-0.05,0.1) in node [left] at (-0.35,1.1);
		\spy on (0.05,-0.305) in node [right] at (0.35,1.1);     
		\end{scope}
		\end{tikzpicture} 
		\vspace{-0.16cm}\\
		
		\rowname{\ac{DTV}} & 
		\begin{tikzpicture}
		\begin{scope}[spy using outlines={rectangle,yellow,magnification=2,size=7mm,connect spies}]
		\node {\includegraphics[height=\tempdima]{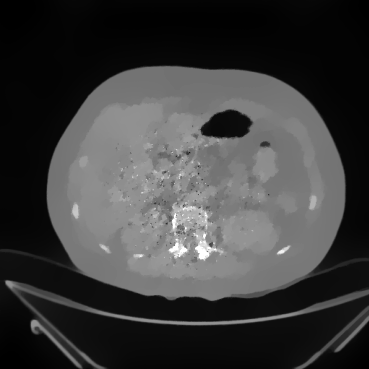}};
		\spy on (-0.05,0.1) in node [left] at (-0.35,1.1);
		\spy on (0.05,-0.305) in node [right] at (0.35,1.1);  
		\end{scope}
		\end{tikzpicture}
		&
		\begin{tikzpicture}
		\begin{scope}[spy using outlines={rectangle,yellow,magnification=2,size=7mm,connect spies}]
		\node {\includegraphics[height=\tempdima]{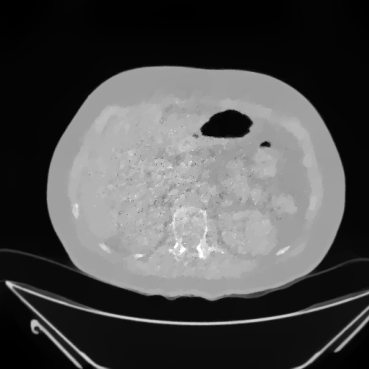}};
		\spy on (-0.05,0.1) in node [left] at (-0.35,1.1);
		\spy on (0.05,-0.305) in node [right] at (0.35,1.1);     
		\end{scope}
		\end{tikzpicture} 
		&
		\begin{tikzpicture}
		\begin{scope}[spy using outlines={rectangle,yellow,magnification=2,size=7mm,connect spies}]
		\node {\includegraphics[height=\tempdima]{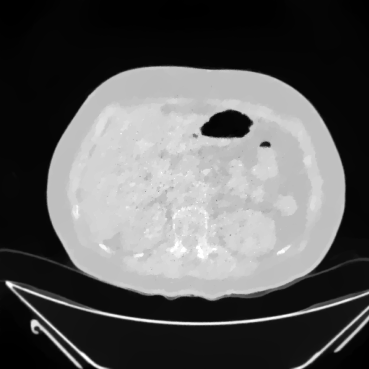}};
		\spy on (-0.05,0.1) in node [left] at (-0.35,1.1);
		\spy on (0.05,-0.305) in node [right] at (0.35,1.1);     
		\end{scope}
		\end{tikzpicture} 
		\vspace{-0.16cm}\\
		
		\rowname{\ac{JTV}} & 
		\begin{tikzpicture}
		\begin{scope}[spy using outlines={rectangle,yellow,magnification=2,size=7mm,connect spies}]
		\node {\includegraphics[height=\tempdima]{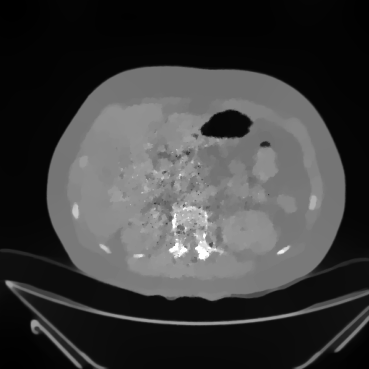}};
		\spy on (-0.05,0.1) in node [left] at (-0.35,1.1);
		\spy on (0.05,-0.305) in node [right] at (0.35,1.1);   
		\end{scope}
		\end{tikzpicture}
		&
		\begin{tikzpicture}
		\begin{scope}[spy using outlines={rectangle,yellow,magnification=2,size=7mm,connect spies}]
		\node {\includegraphics[height=\tempdima]{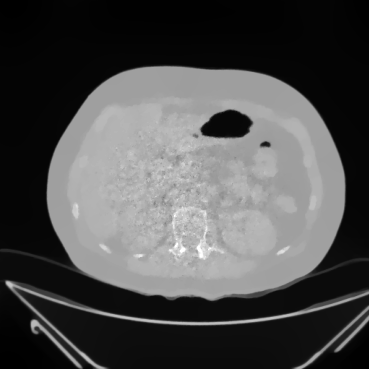}};
		\spy on (-0.05,0.1) in node [left] at (-0.35,1.1);
		\spy on (0.05,-0.305) in node [right] at (0.35,1.1);     
		\end{scope}
		\end{tikzpicture} 
		&
		\begin{tikzpicture}
		\begin{scope}[spy using outlines={rectangle,yellow,magnification=2,size=7mm,connect spies}]
		\node {\includegraphics[height=\tempdima]{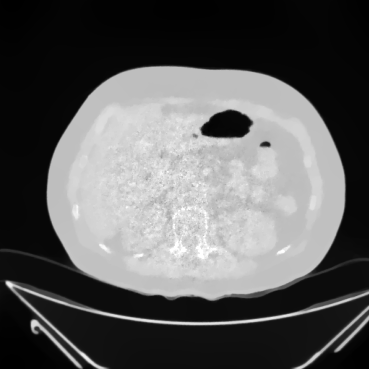}};
		\spy on (-0.05,0.1) in node [left] at (-0.35,1.1);
		\spy on (0.05,-0.305) in node [right] at (0.35,1.1);     
		\end{scope}
		\end{tikzpicture} 
		\vspace{-0.16cm}\\
		
		\rowname{\ac{ASSIST}} & 
		\begin{tikzpicture}
		\begin{scope}[spy using outlines={rectangle,yellow,magnification=2,size=7mm,connect spies}]
		\node {\includegraphics[height=\tempdima]{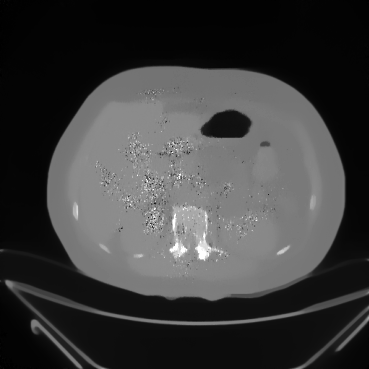}};
		\spy on (-0.05,0.1) in node [left] at (-0.35,1.1);
		\spy on (0.05,-0.305) in node [right] at (0.35,1.1);  
		\end{scope}
		\end{tikzpicture}
		&
		\begin{tikzpicture}
		\begin{scope}[spy using outlines={rectangle,yellow,magnification=2,size=7mm,connect spies}]
		\node {\includegraphics[height=\tempdima]{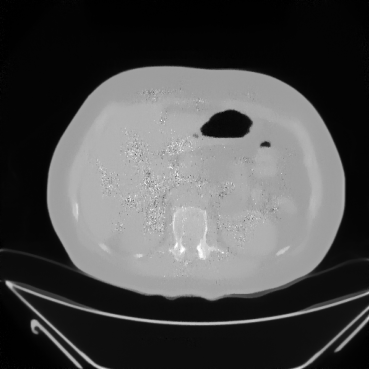}};
		\spy on (-0.05,0.1) in node [left] at (-0.35,1.1);
		\spy on (0.05,-0.305) in node [right] at (0.35,1.1); 
		\end{scope}
		\end{tikzpicture} 
		&
		\begin{tikzpicture}
		\begin{scope}[spy using outlines={rectangle,yellow,magnification=2,size=7mm,connect spies}]
		\node {\includegraphics[height=\tempdima]{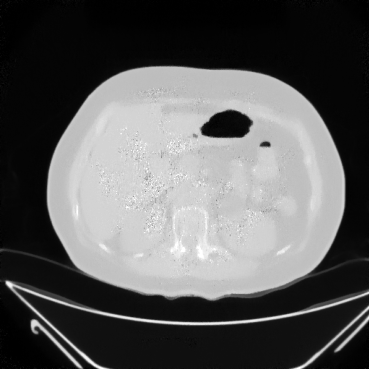}};
		\spy on (-0.05,0.1) in node [left] at (-0.35,1.1);
		\spy on (0.05,-0.305) in node [right] at (0.35,1.1);    
		\end{scope}
		\end{tikzpicture} 
		\vspace{-0.16cm}\\

        \rowname{\ac{RED-CNN}} & 
		\begin{tikzpicture}
		\begin{scope}[spy using outlines={rectangle,yellow,magnification=2,size=7mm,connect spies}]
		\node {\includegraphics[height=\tempdima]{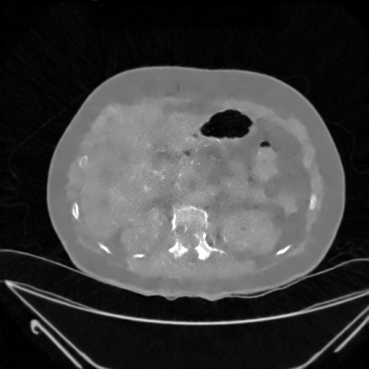}};
		\spy on (-0.05,0.1) in node [left] at (-0.35,1.1);
		\spy on (0.05,-0.305) in node [right] at (0.35,1.1);  
		\end{scope}
		\end{tikzpicture}
		&
		\begin{tikzpicture}
		\begin{scope}[spy using outlines={rectangle,yellow,magnification=2,size=7mm,connect spies}]
		\node {\includegraphics[height=\tempdima]{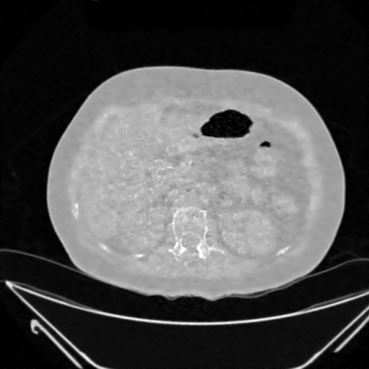}};
		\spy on (-0.05,0.1) in node [left] at (-0.35,1.1);
		\spy on (0.05,-0.305) in node [right] at (0.35,1.1); 
		\end{scope}
		\end{tikzpicture} 
		&
		\begin{tikzpicture}
		\begin{scope}[spy using outlines={rectangle,yellow,magnification=2,size=7mm,connect spies}]
		\node {\includegraphics[height=\tempdima]{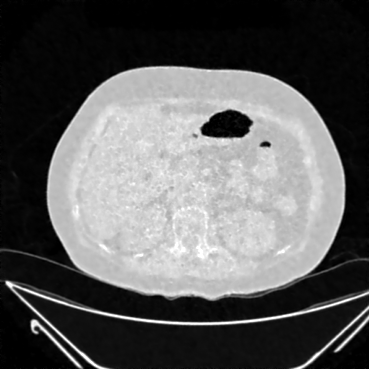}};
		\spy on (-0.05,0.1) in node [left] at (-0.35,1.1);
		\spy on (0.05,-0.305) in node [right] at (0.35,1.1);    
		\end{scope}
		\end{tikzpicture} 
		\vspace{-0.16cm}\\
		
		\rowname{Uconnect} & 
		\begin{tikzpicture}
		\begin{scope}[spy using outlines={rectangle,yellow,magnification=2,size=7mm,connect spies}]
		\node {\includegraphics[height=\tempdima]{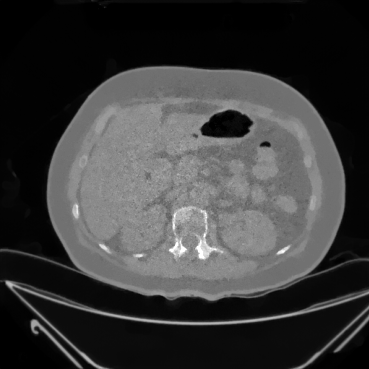}};
		\spy on (-0.05,0.1) in node [left] at (-0.35,1.1);
		\spy on (0.05,-0.305) in node [right] at (0.35,1.1);  
		\end{scope}
		\end{tikzpicture}
		&
		\begin{tikzpicture}
		\begin{scope}[spy using outlines={rectangle,yellow,magnification=2,size=7mm,connect spies}]
		\node {\includegraphics[height=\tempdima]{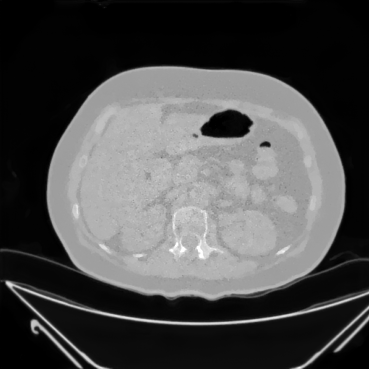}};
		\spy on (-0.05,0.1) in node [left] at (-0.35,1.1);
		\spy on (0.05,-0.305) in node [right] at (0.35,1.1); 
		\end{scope}
		\end{tikzpicture} 
		&
		\begin{tikzpicture}
		\begin{scope}[spy using outlines={rectangle,yellow,magnification=2,size=7mm,connect spies}]
		\node {\includegraphics[height=\tempdima]{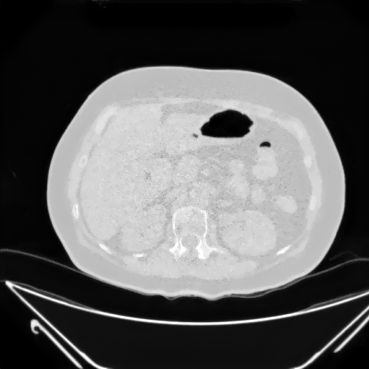}};
		\spy on (-0.05,0.1) in node [left] at (-0.35,1.1);
		\spy on (0.05,-0.305) in node [right] at (0.35,1.1);    
		\end{scope}
		\end{tikzpicture} 
		\vspace{-2.0cm}
		\\
		\rowname{} & 
		\begin{tikzpicture}
		\node {\includegraphics[width=\tempdima]{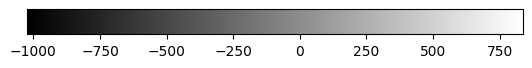}};
		\end{tikzpicture}
		&
		\begin{tikzpicture}
		\node {\includegraphics[width=\tempdima]{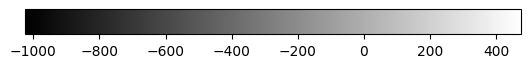}};
		\end{tikzpicture} 
		&
		\begin{tikzpicture}
		\node {\includegraphics[width=\tempdima]{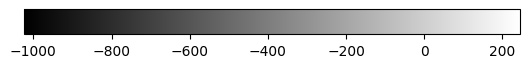}};
		\end{tikzpicture}

	\end{tabular}	
	\caption{\Ac{GT}, noisy images, and denoised images at $E_1=40$, $E_2=80$ and $E_3=140$~keV using different methods.}\label{fig:realdatarecon}
\end{figure}

\begin{figure}[!htb]
	\centering
	\ref{named3} \\
	\subfigure[40~keV]{
		\begin{minipage}[t]{0.5\linewidth}
			\begin{tikzpicture}
			\begin{axis}[
			height=0.7\linewidth,
			width=1.1\linewidth,
			tick label style={font=\scriptsize},
			x label style={at={(axis description cs:0.5,0.2)},font=\scriptsize},
			y label style={at={(axis description cs:0.2,0.5)},anchor=south,font=\scriptsize},
			xlabel={$\beta/\beta_\mathrm{max}$},
			ylabel={PSNR},
			legend columns=3,
			legend entries={Uconnect,\ac{ASSIST},\ac{JTV},\ac{DTV},Huber,\ac{RED-CNN}},
			legend to name=named3,
			]
			\addplot[color=blue, mark=otimes*, mark size=1.5pt] table[x=beta/max_beta, y=PSNR, col sep=comma] {./realdata_psnr_Uconnect_40keV.txt};
            \addlegendentry{Uconnect}
			\addplot[color=violet, mark=diamond*, mark size=1.5pt] table[x=beta/max_beta, y=PSNR, col sep=comma] {./realdata_psnr_ASSIST_40keV.txt};
            \addlegendentry{\ac{ASSIST}}
			\addplot[color=red, mark=diamond*, mark size=1.5pt] table[x=beta/max_beta, y=PSNR, col sep=comma] {./realdata_psnr_JTV_40keV.txt};
            \addlegendentry{\ac{JTV}}
			\addplot[color=green, mark=diamond*, mark size=1.5pt] table[x=beta/max_beta, y=PSNR, col sep=comma] {./realdata_psnr_dTV_40keV.txt};
            \addlegendentry{\ac{DTV}}
			\addplot[color=orange, mark=diamond*, mark size=1.5pt] table[x=beta/max_beta, y=PSNR, col sep=comma] {./realdata_psnr_Huber_40keV.txt};
            \addlegendentry{Huber}
            \addplot[mark=none, black, dashed] coordinates {(0.1,37.53226739208629) (1.0,37.53226739208629)};
            \addlegendentry{\ac{RED-CNN}}
			\end{axis}
			\end{tikzpicture}
		\end{minipage}%
	}%
	\subfigure[60~keV]{
		\begin{minipage}[t]{0.5\linewidth}
			\begin{tikzpicture}
			\begin{axis}[
			height=0.7\linewidth,
			width=1.1\linewidth,
			tick label style={font=\scriptsize},
			x label style={at={(axis description cs:0.5,0.2)},font=\scriptsize},
			xlabel={$\beta/\beta_\mathrm{max}$},
			]
			\addplot[color=blue, mark=otimes*, mark size=1.5pt] table[x=beta/max_beta, y=PSNR, col sep=comma] {./realdata_psnr_Uconnect_60keV.txt};
			\addplot[color=violet, mark=diamond*, mark size=1.5pt] table[x=beta/max_beta, y=PSNR, col sep=comma] {./realdata_psnr_ASSIST_60keV.txt};
			\addplot[color=red, mark=diamond*, mark size=1.5pt] table[x=beta/max_beta, y=PSNR, col sep=comma] {./realdata_psnr_JTV_60keV.txt};
			\addplot[color=green, mark=diamond*, mark size=1.5pt] table[x=beta/max_beta, y=PSNR, col sep=comma] {./realdata_psnr_dTV_60keV.txt};
			\addplot[color=orange, mark=diamond*, mark size=1.5pt] table[x=beta/max_beta, y=PSNR, col sep=comma] {./realdata_psnr_Huber_60keV.txt};
            \addplot[mark=none, black, dashed] coordinates {(0.1,37.76785062140114) (1.0,37.76785062140114)};
			\end{axis}
			\end{tikzpicture}
		\end{minipage}%
	}
	\subfigure[80~keV]{
		\begin{minipage}[t]{0.5\linewidth}
			\begin{tikzpicture}
			\begin{axis}[
			height=0.7\linewidth,
			width=1.1\linewidth,
			tick label style={font=\scriptsize},
			x label style={at={(axis description cs:0.5,0.2)},font=\scriptsize},
			y label style={at={(axis description cs:0.2,0.5)},anchor=south,font=\scriptsize},
			xlabel={$\beta/\beta_\mathrm{max}$},
			ylabel={PSNR},
			]
			\addplot[color=blue, mark=otimes*, mark size=1.5pt] table[x=beta/max_beta, y=PSNR, col sep=comma] {./realdata_psnr_Uconnect_80keV.txt};
			\addplot[color=violet, mark=diamond*, mark size=1.5pt] table[x=beta/max_beta, y=PSNR, col sep=comma] {./realdata_psnr_ASSIST_80keV.txt};
			\addplot[color=red, mark=diamond*, mark size=1.5pt] table[x=beta/max_beta, y=PSNR, col sep=comma] {./realdata_psnr_JTV_80keV.txt};
			\addplot[color=green, mark=diamond*, mark size=1.5pt] table[x=beta/max_beta, y=PSNR, col sep=comma] {./realdata_psnr_dTV_80keV.txt};
			\addplot[color=orange, mark=diamond*, mark size=1.5pt] table[x=beta/max_beta, y=PSNR, col sep=comma] {./realdata_psnr_Huber_80keV.txt};
            \addplot[mark=none, black, dashed] coordinates {(0.1,37.390062851126935) (1.0,37.390062851126935)};
			\end{axis}
			\end{tikzpicture}
		\end{minipage}%
	}%
	\subfigure[100~keV]{
		\begin{minipage}[t]{0.5\linewidth}
			\begin{tikzpicture}
			\begin{axis}[
			height=0.7\linewidth,
			width=1.1\linewidth,
			tick label style={font=\scriptsize},
			x label style={at={(axis description cs:0.5,0.2)},font=\scriptsize},
			y label style={at={(axis description cs:0.2,0.5)},anchor=south,font=\scriptsize},
			xlabel={$\beta/\beta_\mathrm{max}$},
			]
			\addplot[color=blue, mark=otimes*, mark size=1.5pt] table[x=beta/max_beta, y=PSNR, col sep=comma] {./realdata_psnr_Uconnect_100keV.txt};
			\addplot[color=violet, mark=diamond*, mark size=1.5pt] table[x=beta/max_beta, y=PSNR, col sep=comma] {./realdata_psnr_ASSIST_100keV.txt};
			\addplot[color=red, mark=diamond*, mark size=1.5pt] table[x=beta/max_beta, y=PSNR, col sep=comma] {./realdata_psnr_JTV_100keV.txt};
			\addplot[color=green, mark=diamond*, mark size=1.5pt] table[x=beta/max_beta, y=PSNR, col sep=comma] {./realdata_psnr_dTV_100keV.txt};
			\addplot[color=orange, mark=diamond*, mark size=1.5pt] table[x=beta/max_beta, y=PSNR, col sep=comma] {./realdata_psnr_Huber_100keV.txt};
            \addplot[mark=none, black, dashed] coordinates {(0.1,37.444116911924525) (1.0,37.444116911924525)};
			\end{axis}
			\end{tikzpicture}
		\end{minipage}%
	}
	\subfigure[120~keV]{
		\begin{minipage}[t]{0.5\linewidth}
			\begin{tikzpicture}
			\begin{axis}[
			height=0.7\linewidth,
			width=1.1\linewidth,
			tick label style={font=\scriptsize},
			x label style={at={(axis description cs:0.5,0.2)},font=\scriptsize},
			y label style={at={(axis description cs:0.2,0.5)},anchor=south,font=\scriptsize},
			xlabel={$\beta/\beta_\mathrm{max}$},
			ylabel={PSNR},
			]
			\addplot[color=blue, mark=otimes*, mark size=1.5pt] table[x=beta/max_beta, y=PSNR, col sep=comma] {./realdata_psnr_Uconnect_120keV.txt};
			\addplot[color=violet, mark=diamond*, mark size=1.5pt] table[x=beta/max_beta, y=PSNR, col sep=comma] {./realdata_psnr_ASSIST_120keV.txt};
			\addplot[color=red, mark=diamond*, mark size=1.5pt] table[x=beta/max_beta, y=PSNR, col sep=comma] {./realdata_psnr_JTV_120keV.txt};
			\addplot[color=green, mark=diamond*, mark size=1.5pt] table[x=beta/max_beta, y=PSNR, col sep=comma] {./realdata_psnr_dTV_120keV.txt};
			\addplot[color=orange, mark=diamond*, mark size=1.5pt] table[x=beta/max_beta, y=PSNR, col sep=comma] {./realdata_psnr_Huber_120keV.txt};
            \addplot[mark=none, black, dashed] coordinates {(0.1,37.39485279038637) (1.0,37.39485279038637)};
			\end{axis}
			\end{tikzpicture}
		\end{minipage}%
	}%
	\subfigure[140~keV]{
		\begin{minipage}[t]{0.5\linewidth}
			\begin{tikzpicture}
			\begin{axis}[
			height=0.7\linewidth,
			width=1.1\linewidth,
			tick label style={font=\scriptsize},
			x label style={at={(axis description cs:0.5,0.2)},font=\scriptsize},
			y label style={at={(axis description cs:0.2,0.5)},anchor=south,font=\scriptsize},
			xlabel={$\beta/\beta_\mathrm{max}$},
			]
			\addplot[color=blue, mark=otimes*, mark size=1.5pt] table[x=beta/max_beta, y=PSNR, col sep=comma] {./realdata_psnr_Uconnect_140keV.txt};
			\addplot[color=violet, mark=diamond*, mark size=1.5pt] table[x=beta/max_beta, y=PSNR, col sep=comma] {./realdata_psnr_ASSIST_140keV.txt};
			\addplot[color=red, mark=diamond*, mark size=1.5pt] table[x=beta/max_beta, y=PSNR, col sep=comma] {./realdata_psnr_JTV_140keV.txt};
			\addplot[color=green, mark=diamond*, mark size=1.5pt] table[x=beta/max_beta, y=PSNR, col sep=comma] {./realdata_psnr_dTV_140keV.txt};
			\addplot[color=orange, mark=diamond*, mark size=1.5pt] table[x=beta/max_beta, y=PSNR, col sep=comma] {./realdata_psnr_Huber_140keV.txt};
            \addplot[mark=none, black, dashed] coordinates {(0.1,37.50668750292544) (1.0,37.50668750292544)};
			\end{axis}
			\end{tikzpicture}
		\end{minipage}%
	}%
	\caption{\Ac{PSNR} values of the denoised images using different methods. As \ac{RED-CNN} is independent of $\beta$ due to the absence of a penalty term, we depict its evaluation values using a horizontal black dashed line.}\label{fig:psnr_realdata}
\end{figure}
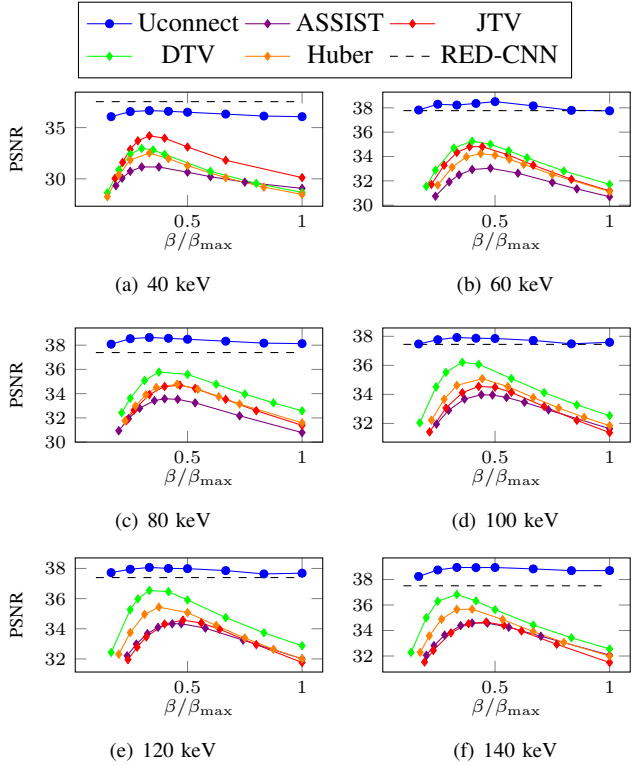

Fig.\@~\ref{fig:realdatarecon} shows the reference images $\boldx_k^\mathrm{patient}$, the noisy images $\boldx_k^\mathrm{noisy}$ and denoised images at 40, 80 and 140~keV with optimal $\beta$ according to the \ac{PSNR} shown in Fig.~\ref{fig:psnr_realdata}. From the noisy images $\boldx_k^\mathrm{noisy}$, we notice that noise is predominantly concentrated in the central regions of the images for all energies, which correspond to structures with high attenuation coefficients. It is noticeable that the results of Huber, \ac{DTV} and \ac{JTV} suffer from severe noise although oversmoothing appears in the periphery of the images. \ac{ASSIST} shows limited capability in distinguishing low-contrast structures especially at 40~keV, and its noise suppression on fine details is inadequate. \ac{RED-CNN} effectively eliminates most noise at 40~keV but fails to preserve the edges of the tissues. Besides, severe noise persists at other energies. On the other hand, the images denoised with Uconnect exhibit the least amount of noise and maintain the highest level of structural and edge preservation across all energies.

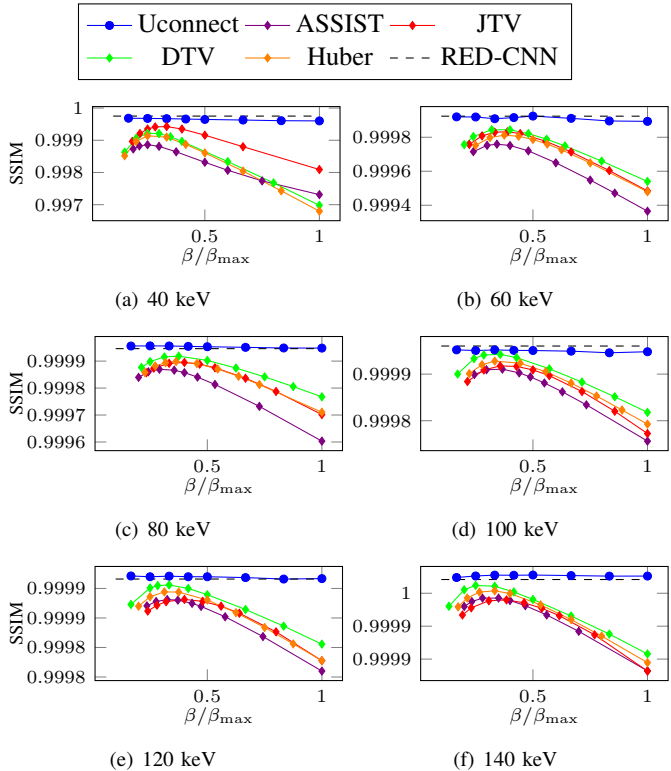
\begin{figure}[!htb]
	\centering
	\ref{named3} \\
	\subfigure[40~keV]{
		\begin{minipage}[t]{0.5\linewidth}
			\begin{tikzpicture}
			\begin{axis}[
			height=0.7\linewidth,
			width=1.1\linewidth,
			tick label style={font=\scriptsize},
			x label style={at={(axis description cs:0.5,0.2)},font=\scriptsize},
			y label style={at={(axis description cs:0.13,0.5)},anchor=south,font=\scriptsize},
			xlabel={$\beta/\beta_\mathrm{max}$},
			ylabel={SSIM},
			scaled y ticks=false,
			y tick label style={/pgf/number format/.cd, fixed, precision=4},
			legend columns=3,
   			legend entries={Uconnect,\ac{ASSIST},\ac{JTV},\ac{DTV},Huber,\ac{RED-CNN}},
			legend to name=named3,
			]
			\addplot[color=blue, mark=otimes*, mark size=1.5pt] table[x=beta/max_beta, y=SSIM, col sep=comma] {./realdata_ssim_Uconnect_40keV.txt};
            \addlegendentry{Uconnect}
			\addplot[color=violet, mark=diamond*, mark size=1.5pt] table[x=beta/max_beta, y=SSIM, col sep=comma] {./realdata_ssim_ASSIST_40keV.txt};
            \addlegendentry{\ac{ASSIST}}
			\addplot[color=red, mark=diamond*, mark size=1.5pt] table[x=beta/max_beta, y=SSIM, col sep=comma] {./realdata_ssim_JTV_40keV.txt};
            \addlegendentry{\ac{JTV}}
			\addplot[color=green, mark=diamond*, mark size=1.5pt] table[x=beta/max_beta, y=SSIM, col sep=comma] {./realdata_ssim_dTV_40keV.txt};
            \addlegendentry{\ac{DTV}}
			\addplot[color=orange, mark=diamond*, mark size=1.5pt] table[x=beta/max_beta, y=SSIM, col sep=comma] {./realdata_ssim_Huber_40keV.txt};
            \addlegendentry{Huber}
            \addplot[mark=none, black, dashed] coordinates {(0.1,0.999750588242089) (1.0,0.999750588242089)};
            \addlegendentry{\ac{RED-CNN}}
			\end{axis}
			\end{tikzpicture}
		\end{minipage}%
	}%
	\subfigure[60~keV]{
		\begin{minipage}[t]{0.5\linewidth}
			\begin{tikzpicture}
			\begin{axis}[
			height=0.7\linewidth,
			width=1.1\linewidth,
			tick label style={font=\scriptsize},
			x label style={at={(axis description cs:0.5,0.2)},font=\scriptsize},
			xlabel={$\beta/\beta_\mathrm{max}$},
			scaled y ticks=false,
			y tick label style={/pgf/number format/.cd, fixed, precision=4},
			]
			\addplot[color=blue, mark=otimes*, mark size=1.5pt] table[x=beta/max_beta, y=SSIM, col sep=comma] {./realdata_ssim_Uconnect_60keV.txt};
			\addplot[color=violet, mark=diamond*, mark size=1.5pt] table[x=beta/max_beta, y=SSIM, col sep=comma] {./realdata_ssim_ASSIST_60keV.txt};
			\addplot[color=red, mark=diamond*, mark size=1.5pt] table[x=beta/max_beta, y=SSIM, col sep=comma] {./realdata_ssim_JTV_60keV.txt};
			\addplot[color=green, mark=diamond*, mark size=1.5pt] table[x=beta/max_beta, y=SSIM, col sep=comma] {./realdata_ssim_dTV_60keV.txt};
			\addplot[color=orange, mark=diamond*, mark size=1.5pt] table[x=beta/max_beta, y=SSIM, col sep=comma] {./realdata_ssim_Huber_60keV.txt};
            \addplot[mark=none, black, dashed] coordinates {(0.1,0.9999241582169551) (1.0,0.9999241582169551)};
			\end{axis}
			\end{tikzpicture}
		\end{minipage}%
	}
	\subfigure[80~keV]{
		\begin{minipage}[t]{0.5\linewidth}
			\begin{tikzpicture}
			\begin{axis}[
			height=0.7\linewidth,
			width=1.1\linewidth,
			tick label style={font=\scriptsize},
			x label style={at={(axis description cs:0.5,0.2)},font=\scriptsize},
			y label style={at={(axis description cs:0.12,0.5)},anchor=south,font=\scriptsize},
			xlabel={$\beta/\beta_\mathrm{max}$},
			ylabel={SSIM},
			scaled y ticks=false,
			y tick label style={/pgf/number format/.cd, fixed, precision=4},
			]
			\addplot[color=blue, mark=otimes*, mark size=1.5pt] table[x=beta/max_beta, y=SSIM, col sep=comma] {./realdata_ssim_Uconnect_80keV.txt};
			\addplot[color=violet, mark=diamond*, mark size=1.5pt] table[x=beta/max_beta, y=SSIM, col sep=comma] {./realdata_ssim_ASSIST_80keV.txt};
			\addplot[color=red, mark=diamond*, mark size=1.5pt] table[x=beta/max_beta, y=SSIM, col sep=comma] {./realdata_ssim_JTV_80keV.txt};
			\addplot[color=green, mark=diamond*, mark size=1.5pt] table[x=beta/max_beta, y=SSIM, col sep=comma] {./realdata_ssim_dTV_80keV.txt};
			\addplot[color=orange, mark=diamond*, mark size=1.5pt] table[x=beta/max_beta, y=SSIM, col sep=comma] {./realdata_ssim_Huber_80keV.txt};
            \addplot[mark=none, black, dashed] coordinates {(0.1,0.9999461520396641) (1.0,0.9999461520396641)};
			\end{axis}
			\end{tikzpicture}
		\end{minipage}%
	}%
	\subfigure[100~keV]{
		\begin{minipage}[t]{0.5\linewidth}
			\begin{tikzpicture}
			\begin{axis}[
			height=0.7\linewidth,
			width=1.1\linewidth,
			tick label style={font=\scriptsize},
			x label style={at={(axis description cs:0.5,0.2)},font=\scriptsize},
			y label style={at={(axis description cs:0.2,0.5)},anchor=south,font=\scriptsize},
			xlabel={$\beta/\beta_\mathrm{max}$},
			scaled y ticks=false,
			y tick label style={/pgf/number format/.cd, fixed, precision=4},
			]
			\addplot[color=blue, mark=otimes*, mark size=1.5pt] table[x=beta/max_beta, y=SSIM, col sep=comma] {./realdata_ssim_Uconnect_100keV.txt};
			\addplot[color=violet, mark=diamond*, mark size=1.5pt] table[x=beta/max_beta, y=SSIM, col sep=comma] {./realdata_ssim_ASSIST_100keV.txt};
			\addplot[color=red, mark=diamond*, mark size=1.5pt] table[x=beta/max_beta, y=SSIM, col sep=comma] {./realdata_ssim_JTV_100keV.txt};
			\addplot[color=green, mark=diamond*, mark size=1.5pt] table[x=beta/max_beta, y=SSIM, col sep=comma] {./realdata_ssim_dTV_100keV.txt};
			\addplot[color=orange, mark=diamond*, mark size=1.5pt] table[x=beta/max_beta, y=SSIM, col sep=comma] {./realdata_ssim_Huber_100keV.txt};
            \addplot[mark=none, black, dashed] coordinates {(0.1,0.9999602649884243) (1.0,0.9999602649884243)};
			\end{axis}
			\end{tikzpicture}
		\end{minipage}%
	}
	\subfigure[120~keV]{
		\begin{minipage}[t]{0.5\linewidth}
			\begin{tikzpicture}
			\begin{axis}[
			height=0.7\linewidth,
			width=1.1\linewidth,
			tick label style={font=\scriptsize},
			x label style={at={(axis description cs:0.5,0.2)},font=\scriptsize},
			y label style={at={(axis description cs:0.12,0.5)},anchor=south,font=\scriptsize},
			xlabel={$\beta/\beta_\mathrm{max}$},
			ylabel={SSIM},
			scaled y ticks=false,
			y tick label style={/pgf/number format/.cd, fixed, precision=4},
			]
			\addplot[color=blue, mark=otimes*, mark size=1.5pt] table[x=beta/max_beta, y=SSIM, col sep=comma] {./realdata_ssim_Uconnect_120keV.txt};
			\addplot[color=violet, mark=diamond*, mark size=1.5pt] table[x=beta/max_beta, y=SSIM, col sep=comma] {./realdata_ssim_ASSIST_120keV.txt};
			\addplot[color=red, mark=diamond*, mark size=1.5pt] table[x=beta/max_beta, y=SSIM, col sep=comma] {./realdata_ssim_JTV_120keV.txt};
			\addplot[color=green, mark=diamond*, mark size=1.5pt] table[x=beta/max_beta, y=SSIM, col sep=comma] {./realdata_ssim_dTV_120keV.txt};
			\addplot[color=orange, mark=diamond*, mark size=1.5pt] table[x=beta/max_beta, y=SSIM, col sep=comma] {./realdata_ssim_Huber_120keV.txt};
            \addplot[mark=none, black, dashed] coordinates {(0.1,0.9999657153161431) (1.0,0.9999657153161431)};
			\end{axis}
			\end{tikzpicture}
		\end{minipage}%
	}%
	\subfigure[140~keV]{
		\begin{minipage}[t]{0.5\linewidth}
			\begin{tikzpicture}
			\begin{axis}[
			height=0.7\linewidth,
			width=1.1\linewidth,
			tick label style={font=\scriptsize},
			x label style={at={(axis description cs:0.5,0.2)},font=\scriptsize},
			y label style={at={(axis description cs:0.13,0.5)},anchor=south,font=\scriptsize},
			xlabel={$\beta/\beta_\mathrm{max}$},
			scaled y ticks=false,
			y tick label style={/pgf/number format/.cd, fixed, precision=4},
			]
			\addplot[color=blue, mark=otimes*, mark size=1.5pt] table[x=beta/max_beta, y=SSIM, col sep=comma] {./realdata_ssim_Uconnect_140keV.txt};
			\addplot[color=violet, mark=diamond*, mark size=1.5pt] table[x=beta/max_beta, y=SSIM, col sep=comma] {./realdata_ssim_ASSIST_140keV.txt};
			\addplot[color=red, mark=diamond*, mark size=1.5pt] table[x=beta/max_beta, y=SSIM, col sep=comma] {./realdata_ssim_JTV_140keV.txt};
			\addplot[color=green, mark=diamond*, mark size=1.5pt] table[x=beta/max_beta, y=SSIM, col sep=comma] {./realdata_ssim_dTV_140keV.txt};
			\addplot[color=orange, mark=diamond*, mark size=1.5pt] table[x=beta/max_beta, y=SSIM, col sep=comma] {./realdata_ssim_Huber_140keV.txt};
            \addplot[mark=none, black, dashed] coordinates {(0.1,0.9999709594036629) (1.0,0.9999709594036629)};
			\end{axis}
			\end{tikzpicture}
		\end{minipage}%
	}%
	\caption{\Ac{SSIM} values of the denoised images using different methods. As \ac{RED-CNN} is independent of $\beta$ due to the absence of a penalty term, we depict its evaluation values using a horizontal black dashed line.}\label{fig:ssim_realdata}
\end{figure}

We carried out the same evaluations, \ie{} \ac{PSNR} and \ac{SSIM}, as in Section~\ref{sec:img}. Fig.\@~\ref{fig:psnr_realdata} and Fig.\@~\ref{fig:ssim_realdata} respectively show the \ac{PSNR} and \ac{SSIM} values (using  $\boldx_k^\mathrm{patient}$, $k=1,\dots,6$, as reference) at each of the  6 energy bins for each method for a range of $\beta$-values. Uconnect exhibits significantly superior \ac{PSNR} and \ac{SSIM} values compared to Huber, \ac{DTV}, \ac{JTV}, and \ac{ASSIST}. \ac{RED-CNN} and Uconnect exhibit comparable \ac{SSIM} values across all energy levels. In terms of \ac{PSNR}, while \ac{RED-CNN} shows a higher value than Uconnect at 40~keV, Uconnect surpasses \ac{RED-CNN} at other energy levels. Taking into account the qualitative comparison illustrated in Fig.\@~\ref{fig:realdatarecon}, it becomes evident that Uconnect outperforms all other methods considering all energies.

%% file: discussion.tex
\section{Discussion}\label{sec:discussion}

We proposed a novel method that takes the form of a trained penalty to solve the problem of synergistic reconstruction for spectral \ac{CT}. In this paper, we used the pretrained U-Nets to ``connect'' images at all energies to a reference $\boldz$, so that the training can be supervised. We considered as reference the image at the lowest energy. Our U-Net based regularization, referred to as Uconnect, showed enhanced performance for both image reconstruction and image denoising tasks. The mappings $\left\{\boldf_k\right\}$ provide regularization  (i)  by their training, as they are trained on ``clean'' images (as initially suggested in \cite{gong2018iterative}), and (ii) by sharing the information across channels, in a similar fashion as \ac{TDiL} but without patch extraction. Note that our method can be generalized to a single architecture with $K$ outputs, each of which corresponding to an energy bin, in order to further leverage the inter-energy information during training.

As pointed out in Section~\ref{sec:model}, the pretrained mappings $\left\{\boldf_k \right\}$ do not exactly reflect the physics, as $\boldz$ represents an attenuation image. Alternatively,  models such as \ac{WGAN} or \ac{VAE}, where the latent space is ``trained'' to encode the physical properties of the patient, should be used \cite{duff2021regularising}. These models have been applied in \ac{PET}/\ac{CT} \cite{pinton2023synergistic} and \ac{PET}/\ac{MRI} \cite{gautier2023vae}. Similar models may also be used to generate material images and could be used for one-step material decomposition. However unsupervised training with a large number of channels remains challenging.

The results of the proposed approach can be potentially improved by fine-tuning all parameters, including $\alpha$ and $\left\{\gamma_k\right\}$ in \eqref{eq:r-final}, which we have not yet investigated. Additionally, we can explore other regularizers for the reference $\boldz$ in our method, such as by replacing Huber with \ac{TV} or \ac{DTV}. 

An major limit to our validation is that our phantom data were based on \acp{VMI}, which correspond to an oversimplified setting. In addition, \acp{VMI} are artificially correlated, which may bias the results. Further validation should include real raw \ac{PCCT} projection data and models trained from \ac{PCCT}  binned images.

Finally, our comparison can be expanded beyond other synergistic reconstruction methods, such as \ac{TDiL} \cite{zhang2016tensor} and SOUL-Net \cite{chen2022soul}, a recent work using an unrolling architecture. 

%% file: conclusion.tex
\section{Conclusion}\label{sec:conclusion}

In this work, we used a \ac{CNN} to derive a novel synergistic penalty term for spectral \ac{CT} reconstruction. The proposed approach is a combination on the \ac{CNN} representation technique \cite{gong2018iterative} and multichannel dictionary learning \cite{zhang2016tensor}. 

The presented results demonstrate that Uconnect has the ability to reduce noise while preserving structures. Both visual inspection and quantitative measurements show significant improvements for our proposed method in synergistic spectral \ac{CT} reconstruction. While there is certainly room for improvement, we believe that the proposed methodological framework provides a solid foundation for future research in this field. Further refinement of our approach may be achieved by unrolling our architecture, a development that warrants further investigation.